\documentclass[aps,prd,twocolumn,nofootinbib]{revtex4-1}
\usepackage{epsfig}
\usepackage[colorlinks,linkcolor=blue,anchorcolor=blue,citecolor=blue,urlcolor=blue,breaklinks=true]{hyperref}
\usepackage{graphics}
\usepackage{slashed}
\usepackage{color}
\usepackage{amsmath}
\usepackage{bm}
\usepackage{setspace}
\usepackage{caption}
\usepackage{multirow}
\usepackage{subfigure}
\begin{document}
\author{Cheng-Ming Li$^{1}$}\email{licm@zzu.edu.cn}
\author{Shu-Yu Zuo$^{2}$}\email{zuoshuyu12@163.com}
\author{Yan Yan$^{3}$}\email{2919ywhhxh@163.com}
\author{Ya-Peng Zhao$^{4}$}\email{zhaoyapeng2013@hotmail.com}
\author{Fei Wang$^{1}$}\email{feiwang@zzu.edu.cn}
\author{Yong-Feng Huang$^{5}$}\email{hyf@nju.edu.cn}
\author{Hong-Shi Zong$^{6,7,8}$}\email{zonghs@nju.edu.cn}
\address{$^{1}$ School of Physics and Microelectronics, Zhengzhou University, Zhengzhou 450001, China}
\address{$^{2}$ College of Science, Henan University of Technology, Zhengzhou 450000, China}
\address{$^{3}$ School of mathematics and physics, Changzhou University, Changzhou, Jiangsu 213164, China}
\address{$^{4}$ Collage of Physics and Electrical Engineering, Anyang Normal University, Anyang, 455000, China}
\address{$^{5}$ School of Astronomy and Space Science, Nanjing University, Nanjing 210023, China}
\address{$^{6}$ Department of Physics, Nanjing University, Nanjing 210093, China}
\address{$^{7}$ Department of Physics, Anhui Normal University, Wuhu, Anhui 241000, China}
\address{$^{8}$ Nanjing Institute of Proton Source Technology, Nanjing, 210046 China}
\title{Strange quark stars within proper time regularized (2+1)-flavor NJL model}

\begin{abstract}
In this work we use the equation of state (EOS) of (2+1)-flavor Nambu-Jona-Lasinio (NJL) model to study the structure of the strange quark star. With a new free parameter $\alpha$, the Lagrangian is constructed by two parts, the original NJL Lagrangian and the Fierz transformation of it, as $\mathcal{L}=(1-\alpha)\mathcal{L}_{NJL}+\alpha\mathcal{L}_{Fierz}$. To determine the range of $\alpha$, we compare the binding energies in the 2-flavor and (2+1)-flavor cases. We also consider the constraints of chemical equilibrium and electric charge neutrality in the strange quark star and choose six representative EOSs with different $\alpha$ and $B$ (bag constant) to study their influence on the structure of the strange quark star. As a result, we find that a larger $\alpha$ and a smaller $B$ corresponds to a heavier star with a stiffer EOS. Furthermore, the heaviest strange quark star is in agreement with not only the recent mass observation of PSR J0740+6620 and the X-ray observations on radius measurements, but also the constraint on tidal deformability of GW170817.

\bigskip

\bigskip


\end{abstract}


\maketitle

\section{INTRODUCTION}
The equation of state (EOS) plays a critical role in the study of the neutron star. Substituting an EOS into the Tolman-Oppenheimer-Volkoff (TOV) equation, we can get the corresponding mass-radius relation of the star. In general, the EOS should meet many constraints, such as the mass from pulsar observations, the radius measurement from X-ray observations, and the tidal deformability from gravitational wave (GW) observations. On one hand, the pulsar mass measured in the recent astronomical observation PSR J0740+6620~\cite{Cromartie2019}, 2.14$^{+0.10}_{-0.09}$ $M_{\odot}$ (solar mass), has become the most massive one till now, even larger than that of PSR J0348+0432~\cite{Antoniadis1233232} with 2.01$\pm$0.04 $M_{\odot}$, excluding many soft EOSs that can not produce so massive star. And the X-ray observations especially the Neutron Star Interior Composition Explorer (NICER) X-ray timing observations are supplying more and more precise measurements recently~\cite{Bogdanov_2019,Capano:2019eae,Riley_2019}, for example, in Ref.~\cite{Capano:2019eae}, the radius is limited to $11.0^{+0.9}_{-0.6}$ km for the neutron star with $1.4 M_{\odot}$. On the other hand, the GW observation GW170817 during the binary neutron star (BNS) merger gives a constraint on the tidal deformability of the star for the low-spin priors, estimated to be $\Lambda(1.4M_{\odot})\leq800$ and $\tilde{\Lambda}\leq800$ in the early work~\cite{PhysRevLett.119.161101}, and revised to be $\tilde{\Lambda}\sim340^{+580}_{-240}$ for the case of symmetric $90\%$ credible interval and $\tilde{\Lambda}\sim340^{+490}_{-290}$ for the case of highest posterior density (HPD) $90\%$ credible interval based on the waveform model TaylorF2 in the recent study~\cite{PhysRevX.9.011001}, thus will exclude many stiff EOSs with large tidal deformabilities.

Considering that the neutron star is composed of strong interacted dense matter in relatively low temperature, it is imperative to study the EOS and the structure of the neutron star in the framework of the quantum chromodynamics (QCD). It is known that the QCD has two important properties, the color confinement and dynamical chiral symmetry breaking. At low chemical potential and low temperature, the quarks are confined in hadrons. However, at high chemical potential, the quarks are deconfined, thus the observed pulsar in this case could be a quark star rather than the traditional neutron star. Then a question arises: which kind of quark stars should actually exist, the non-strange quark star only containing $u$, $d$ quarks or the strange quark star containing $u$, $d$ and $s$ quarks? Different models and perspectives give different answers. In Refs~\cite{PhysRevD.30.272,Dexheimer2013}, the 3-flavor system is demonstrated to be more stable than the 2-flavor one, but in Refs.~\cite{doi:10.1143/JPSJ.58.3555,doi:10.1143/JPSJ.58.4388}, the opposite side is right. Recently, a study~\cite{PhysRevLett.120.222001} indicates that the non-strange quark matter can be the ground state of baryonic matter only for baryon number larger than a certain value. In the light of this conclusion, some studies~\cite{PhysRevD.100.043018,PhysRevD.100.123003,Zhang:2019mqb} investigate the possibility of the non-strange quark star and calculate the structure of it. But it should be pointed out that this conclusion of Ref.~\cite{PhysRevLett.120.222001} is actually made by the comparison of the energy per baryon (i.e., the binding energy) between the strange and non-strange quark matter with an effective model. In this work, we will revisit this question and study the possibility that the quark star is a strange one with the strange quark system being more stable.

Theoretically, the matter in the quark star is very dense and interacted so strong that the perturbative QCD is invalid here, and the "sign problem" in the lattice QCD (LQCD) makes it difficult to perform calculations at finite chemical potential. However, some effective models including the
Dyson-Schwinger equations (DSEs)~\cite{ROBERTS1994477,Roberts2000S1,doi:10.1142/S0218301303001326,Cloet20141,PhysRevD.90.114031,PhysRevD.91.034017,*PhysRevD.91.056003}, the quantum electrodynamics in 2+1 dimensions (QED3)~\cite{ROBERTS1994477,PhysRevD.29.2423,PhysRevD.90.036007,PhysRevD.90.073013}, and the Nambu-Jona-Lasinio (NJL) model~\cite{RevModPhys.64.649,Buballa2005205,Cui2013,KOHYAMA2015682,doi:10.1142/S0217751X17500610,PhysRevD.99.076006} are very useful in this scheme. In Refs.~\cite{PhysRevD.84.105023,PhysRevD.91.105002,PhysRevD.92.054012,doi:10.1142/S0217732317500511,PhysRevD.96.043008,PhysRevD.97.023018} and Refs.~\cite{10.1143/PTPS.186.81,Masuda01072013,PhysRevC.93.035807,PhysRevD.94.094001,PhysRevD.95.056018,PhysRevD.97.103013,PhysRevD.98.083013}, the structure of the hybrid star is investigated with the DSEs and NJL model, respectively. Unfortunately, the quark EOSs in these studies are still very soft and can not support a quark star with two solar mass. In a recent study~\cite{Wang2019}, the authors propose a modified NJL model containing both the original model Lagrangian and the Fierz transformation of it with the parameter $(1-\alpha)$ and $\alpha$ adjusting the weight of these two parts, respectively, thus producing stiffer EOSs than before. In this scheme, the non-strange quark star has been studied for the 2-flavor case~\cite{PhysRevD.100.043018,PhysRevD.100.123003}, and the results satisfy both two solar mass constraint from PSR J0348+0432~\cite{Antoniadis1233232} and the tidal deformability constraint $\Lambda(1.4 M_{\odot})\leq800$ from GW170817 in the previous study~\cite{PhysRevLett.119.161101}. In this work, we will extend the modified NJL model mentioned above to the (2+1)-flavor case and give the EOS with the mean field approximation and proper time regularization (PTR). The mass-radius relation and the tidal deformability of the strange quark star with the corresponding EOS are also studied.

The paper is organized as follows. In Sec.~\ref{one}, we introduce the EOS of the strange quark star with the modified (2+1)-flavor NJL model. To determine whether the quark matter is more stable for (2+1)-flavor case than for 2-flavor case, we compare the binding energies in these two cases for different $\alpha$ and $B$. In Sec.~\ref{two}, we use the new (2+1)-flavor EOS to study the structure of the strange quark star, and the corresponding tidal deformability is also studied with different $\alpha$ and $B$. For conclusion, we give a brief summary and discussion in Sec.~\ref{three}.

\section{EOS of the strange quark matter with NJL model}\label{one}
In this section, we give a brief introduction of the modified (2+1)-flavor NJL model, and the EOS is also deduced with PTR and mean-field approximation. In general, the Lagrangian of (2+1)-flavor NJL model has the following form,
\begin{eqnarray}
\mathcal{L}_{\rm NJL}=&&\bar{\psi}(i{\not\!\partial}-m)\psi+\sum^8_{\rm i=0}G[(\bar{\psi}\lambda_{\rm i}\psi)^2+(\bar{\psi}i\gamma^{5}\lambda_{\rm i}\psi)^2]\nonumber\\
&&-K\,({\rm det}[\bar{\psi}(1+\gamma^{5})\psi]+{\rm det}[\bar{\psi}(1-\gamma^{5})\psi]),\,\,\label{lagrangian}
\end{eqnarray}
where $G$ and $K$ are the four-fermion and six-fermion interaction coupling constant, respectively, $\lambda^{\rm i}$ (${\rm i}=1\rightarrow 8$) is the Gell-Mann matrix in flavor space and $\lambda^0=\sqrt{\frac{2}{3}}\,I$ ($I$ is the identity matrix).

As a purely technical device to examine the effect of a rearrangement of fermion field operators, the Fierz transformation of the lagrangian $\mathcal{L}_{\rm NJL}$ is
\begin{eqnarray}
\mathcal{L}_{\rm F}=&&\bar{\psi}(i{\not\!\partial}-m)\psi-\frac{1}{2}\sum^8_{\rm a=0}G[(\bar{\psi}\gamma^{\mu}\lambda_{\rm a}^C\psi)^2-(\bar{\psi}\gamma^{\mu}\gamma^{5}\lambda_{\rm a}^C\psi)^2]\nonumber\\
&&-K\,({\rm det}[\bar{\psi}(1+\gamma^{5})\psi]+{\rm det}[\bar{\psi}(1-\gamma^{5})\psi]),\,\,\label{lagrangianf}
\end{eqnarray}
where $\lambda_{\rm a}^C$ (${\rm a}=0\rightarrow 8$) also has a same definition as $\lambda_{\rm i}$ (${\rm a}=0\rightarrow 8$) in the above but in color space. Actually, the Fierz transformation of the six-fermion term $-K\,({\rm det}[\bar{\psi}(1+\gamma^{5})\psi]+{\rm det}[\bar{\psi}(1-\gamma^{5})\psi])$ will introduce the terms including pairs of color octet quark bilinears as a part~\cite{KLIMT1990386,KLIMT1990429}. And in Ref.~\cite{RevModPhys.64.649}, it is demonstrated that a Fierz transformation of the six-fermion interaction can be defined as that transformation that leaves the interaction invariant under all possible permutations of the quark spinors $\psi$ occurring in it, thus the six-fermion term does not change in Eq.~(\ref{lagrangianf}). On the other hand, even one introduces other four-fermion terms that contains the color octet quark bilinears and are invariant under $\rm{SU}(3)_V\otimes\rm{SU}(3)_A\otimes\rm{U}(1)_V\otimes\rm{U}(1)_A$ in the original Lagrangian~\cite{KLIMT1990386}, the results are qualitatively similar to those obtained neglecting them~\cite{BERNARD1988753}. Therefore, in the following calculation, we will just consider the contribution of color singlet terms for simplicity.

According to Ref.~\cite{10.1143/PTP.74.765}, the general mean field approximation approach without Fierz transformation is demonstrated to be not self-consistent. Thus in this work, we employ a new self-consistent way to deal with it, introducing the weighting factor $(1-\alpha)$ and $\alpha$ to combine the Lagrangian $\mathcal{L}_{\rm NJL}$ and its Fierz transformation $\mathcal{L}_{\rm F}$ linearly,
\begin{eqnarray}
\mathcal{L}=&&(1-\alpha)\mathcal{L}_{NJL}+\alpha\mathcal{L}_{F}\nonumber\\
=&&\bar{\psi}(i{\not\!\partial}-m)\psi+(1-\alpha) G\sum^8_{\rm i=0}[(\bar{\psi}\lambda_{\rm i}\psi)^2+(\bar{\psi}i\gamma^{5}\lambda^{\rm i}\psi)^2]\nonumber\\
&&-\alpha\frac{G}{2}[(\bar{\psi}\gamma^{\mu}\lambda_0^C\psi)^2-(\bar{\psi}\gamma^{\mu}\gamma^{5}\lambda_ 0^C\psi)^2]\nonumber\\
&&-K\,({\rm det}[\bar{\psi}(1+\gamma^{5})\psi]+{\rm det}[\bar{\psi}(1-\gamma^{5})\psi]).\,\,\label{lagrangiantotal}
\end{eqnarray}

Then we take the mean field approximation to obtain the dynamical quark mass $M_{\rm i}$ and the renormalized chemical potential $\mu_{\rm i}^{\prime}$ of flavor i, respectively,
\begin{equation}\label{gapeq}
  M_{\rm i}=m_{\rm i}-4G\langle\bar{\psi}\psi\rangle_{\rm i}+2K\langle\bar{\psi}\psi\rangle_{\rm j}\langle\bar{\psi}\psi\rangle_{\rm k},
\end{equation}
\begin{equation}\label{rechempot}
  \mu_{\rm i}^{\prime}=\mu_{\rm i}-\frac{2\,\alpha}{N_{\rm c}\,(1-\alpha)}G\langle\psi^{+}\psi\rangle_{\rm i}.
\end{equation}
The $\langle\bar{\psi}\psi\rangle_{\rm i}$ and $\langle\psi^{+}\psi\rangle_{\rm i}$ in Eq.~(\ref{gapeq}) and (\ref{rechempot}) are the quark condensate and quark number density of flavor i, respectively. At zero temperature, they are defined as
\begin{eqnarray}
  \langle\bar{\psi}\psi\rangle_{\rm i} &=& -\int\frac{{\rm d}^4p}{(2\pi)^4}{\rm Tr}[iS_{\rm i}(p^2)]\nonumber\\
  &=& -N_{\rm c}\int_{-\infty}^{+\infty}\frac{{\rm d}^4p}{(2\pi)^4}\frac{4iM_{\rm i}}{p^2-M_{\rm i}^2},\,\,\label{qcondensate}
\end{eqnarray}
\begin{eqnarray}
  \langle\psi^+\psi\rangle_{\rm i} &=& -\int\frac{{\rm d}^4p}{(2\pi)^4}{\rm Tr}\left[iS_{\rm i}(p^2)\gamma_0\right]\nonumber\\
   &=& 2N_{\rm c}\int\frac{{\rm d}^3p}{(2\pi)^3}\theta(\mu_{\rm i}^{\prime}-\sqrt{p^2+M_{\rm i}^2}),\label{qnd}
\end{eqnarray}
where the trace ``${\rm Tr}$" is taken in Dirac and color spaces, and $S_{\rm i}(p^2) = \frac{1}{{\not\,p}-M_{\rm i}}$ is the quark propagator of flavor i.

Now we can see that the construction of the Lagrangian $\mathcal{L}$ is equivalent to adding the vector-scalar channel in the four-fermion interaction term of the original Lagrangian $\mathcal{L}_{\rm NJL}$. In addition, from Eq.~(\ref{rechempot}), via the introduction of $\mathcal{L}_{\rm F}$ especially when $\alpha$ is taken as large values ($\alpha\rightarrow1$), the contribution of the Fierz-transformed part can not be neglected, but usually ignored in the mean field approximation when just counting in $\mathcal{L}_{\rm NJL}$. Therefore, from this viewpoint, our approach is different from the original one.

To perform the following calculations, a Wick rotation from Minkowski space to Euclidean space and the proper-time regularization (PTR) are employed. According to the definition, the PTR equals to replacing the ultra-violet divergent integrand $\frac{1}{A^n}$ as an integral of its exponential function, that is,
\begin{eqnarray}
  \frac{1}{A^n} &=& \frac{1}{(n-1)!}\int_{0}^{\infty}{\rm d}\tau\tau^{n-1}e^{-\tau A}\nonumber \\
   & &\xrightarrow{\rm UV cutoff} \frac{1}{(n-1)!}\int_{\tau_{\rm UV}}^{\infty}{\rm d}\tau\tau^{n-1}e^{-\tau A},\,\,\label{regularization}
\end{eqnarray}
where the integral limit $\tau_{\rm UV}$ is related to the ultra-violet cutoff $\Lambda_{\rm UV}$ as $\tau_{\rm UV}=\Lambda_{\rm UV}^{-2}$. In addition, when we extend the calculation from zero chemical potential to the finite case, it is equivalent to introduce a transformation~\cite{PhysRevC.71.015205} that
\begin{equation}\label{muinpfour}
  p_4\rightarrow p_4+i\mu_{\rm i}^{\prime}.
\end{equation}

Then we can derive the quark condensate and quark number density in the following,

(\romannumeral1). for $T=0,\,\mu_{\rm i}^{\prime}=0$,
\begin{eqnarray}
   \langle\bar{\psi}\psi\rangle_{\rm i} &=& -N_{\rm c}\int_{-\infty}^{+\infty}\frac{{\rm d}^4p^{\rm E}}{(2\pi)^4}\frac{4iM_i}{(p^{\rm E})^{2}+M_i^2}\nonumber\\
    &=& -\frac{N_{\rm c}}{(2\pi)^4}\int_{-\infty}^{+\infty}\int_{-\infty}^{+\infty}{\rm d}^3\overrightarrow{p}{\rm d}p_4\frac{4M_i}{p_4^2+\overrightarrow{p}^2+M_i^2}\nonumber \\
  &=& -\frac{3M_i}{\pi^2}\int_{0}^{+\infty}{\rm d}p\frac{p^2}{\sqrt{p^2+M_i^2}}\nonumber\\
   &=& -\frac{3M_i}{\pi^{\frac{2}{5}}}\int_{\tau_{\rm UV}}^{\infty}\int_{0}^{+\infty}{\rm d}\tau {\rm d}p\tau ^{-\frac{1}{2}}p^2e^{-\tau (M_i^2+p^2)}\nonumber\\
    &=& -\frac{3M_i}{4\pi^2}\int_{\tau_{\rm UV}}^{\infty}{\rm d}\tau \frac{e^{-\tau M_i^2}}{\tau^2},\,\,\label{zzqcondensate}
  \end{eqnarray}

(\romannumeral2). for $T=0,\,\mu_{\rm i}^{\prime}\neq0$,
\begin{eqnarray}
  \langle\bar{\psi}\psi\rangle_{\rm i}&=&-N_{\rm c}\int_{-\infty}^{+\infty}\frac{{\rm d}^4p^{\rm E}}{(2\pi)^4}\frac{4iM_i}{(p^{\rm E})^{2}+M_i^2}\nonumber\\
  &=&-N_{\rm c}\int_{-\infty}^{+\infty}\frac{{\rm d}^4p}{(2\pi)^4}\frac{4M_{\rm i}}{(p_4+i\mu_{\rm i}^{\prime})^2+M_{\rm i}^2+\overrightarrow{p}^2}\nonumber\\
   &=&-\frac{3M_{\rm i}}{\pi^3}\int_{0}^{+\infty}{\rm d}p\int_{-\infty}^{+\infty}{\rm d}p_4\frac{p^2}{(p_4+i\mu_{\rm i}^{\prime})^2+M_{\rm i}^2+p^2}\nonumber\\
   &=&\left\{\!
   \begin{small}
  \begin{array}{lcl}
\!\!\displaystyle{-\frac{3M_{\rm i}}{\pi^2}\!\!\!\int_{\!\!\sqrt{{\mu_{\rm i}^{\prime}}^2-M_{\rm i}^2}}^{+\infty}\!{\rm d}p\textstyle{\frac{\left[1-{\rm Erf}(\sqrt{M_{\rm i}^2+p^2}\sqrt{\tau_{\rm UV}})\right]p^2}{\sqrt{M_{\rm i}^2+p^2}}}},\,M_{\rm i}<\mu_{\rm i}^{\prime}\\
\displaystyle{\frac{3M_{\rm i}}{4\pi^2}\left[\textstyle{-M_{\rm i}^2{\rm Ei}(-M_{\rm i}^2\tau_{\rm UV})-\frac{e^{-M_{\rm i}^2\tau_{\rm UV}}}{\tau_{\rm UV}}}\right]},\,\,M_{\rm i}>\mu_{\rm i}^{\prime}
  \end{array}
  \end{small}\right.\nonumber\\
  \label{zfqcondensate}
\end{eqnarray}
\begin{eqnarray}
  \langle\psi^+\psi\rangle_{\rm i} &=& 2N_{\rm c}\int\frac{{\rm d}^3p}{(2\pi)^3}\theta(\mu_{\rm i}^{\prime}-\sqrt{p^2+M_{\rm i}^2})\nonumber\\
  &=& \left\{
\begin{array}{lcl}
 \frac{1}{\pi^2}(\sqrt{{\mu_{\rm i}^{\prime}}^2-M_{\rm i}^2})^3,             & &\mu_{\rm i}^{\prime}>M_{\rm i}\\
  0,                                                & &\mu_{\rm i}^{\prime}<M_{\rm i}
   \end{array}
   \right.\label{qnd}
\end{eqnarray}
where the superscript E represents the Euclidean space. Ei(x)$=-\int_{-x}^{+\infty}{\rm d}y\frac{e^{-y}}{t}$ and Erf(x)$=\frac{2}{\sqrt{\pi}}\int_{0}^{x}e^{-\eta^2}{\rm d}\eta$ are the Exponential Integral function and error function, respectively.

From Eqs.~(\ref{gapeq}) and (\ref{rechempot}), it is noted that the introduction of $\mathcal{L}_{\rm F}$ only contributes to the renormalized chemical potential, but not the gap equation and the dynamical quark mass. Thus at zero temperature and chemical potential, apart from $\alpha$, the parameter fixing work is still same with the original case for $\mathcal{L}_{\rm NJL}$. Similar to the process in Ref.~\cite{PhysRevD.98.083013}, we also fit the parameters ($M_{\rm u}, \Lambda_{\rm UV}$, $M_{\rm s}$, $G$, $K$) to reproduce the experimental data ($f_{\pi}=92$ MeV, $M_{\pi}=135$ MeV, $M_{K^0}=495$ MeV, $M_{\eta}=548$ MeV, $M_{\eta '}=958$ MeV), with a free parameter $m_{\rm u}$ pre-fixed before the fitting. According to the recent Review of Particle Physics~\cite{PhysRevD.98.030001}, the current quark mass $m_{\rm u}$ and $m_{\rm s}$ are predicted to be $\bar{m}=(m_{\rm u}+m_{\rm d})/2=3.5^{+0.5}_{-0.2}$ MeV, $m_{\rm s}=95^{+9}_{-3}$ MeV, respectively\footnote{The exact isospin symmetry between u and d quark is
employed in this work, thus $m_{\rm u} = m_{\rm d} = \bar{m}$.}. Our parameter sets satisfying these constraints on the current quark masses are shown in Table.~\ref{parameters}. The two parameter sets in Table.~\ref{parameters} do not have significant difference, and their corresponding EOSs also turn out to be very similar from Ref.~\cite{PhysRevD.98.083013}. Therefore, we will choose the parameter set with $m_{\rm u}=3.4$ MeV as a representative one to perform the following calculation.
\begin{table}
\caption{Parameter sets satisfying the constraints on the current quark masses $m_{\rm u}$ and $m_{\rm s}$ from the recent Review of Particle Physics~\cite{PhysRevD.98.030001}. The unit of the coupling constants $G$ and $K$ are MeV$^{-2}$ and MeV$^{-5}$, respectively, and the other parameters in this table have the unit of MeV.}\label{parameters}
\begin{tabular}{p{0.6cm} p{0.6cm} p{0.7cm} p{1.9cm} p{1.9cm}p{0.6cm}p{0.6cm}}
\hline\hline
$m_{\rm u}$&$\,m_{\rm s}$&$\Lambda_{\rm UV}$&$\qquad G$&$\qquad K$&$M_{\rm u}$&$M_{\rm s}$\\
\hline
3.3&102&1350$\,\,$&$\,1.46\times10^{-6}$&$2.55\times10^{-14}$&195&361\\
3.4&104&1330$\,\,$&$\,1.51\times10^{-6}$&$2.75\times10^{-14}$&197&364\\
\hline\hline
\end{tabular}
\end{table}
\begin{figure}
\includegraphics[width=0.47\textwidth]{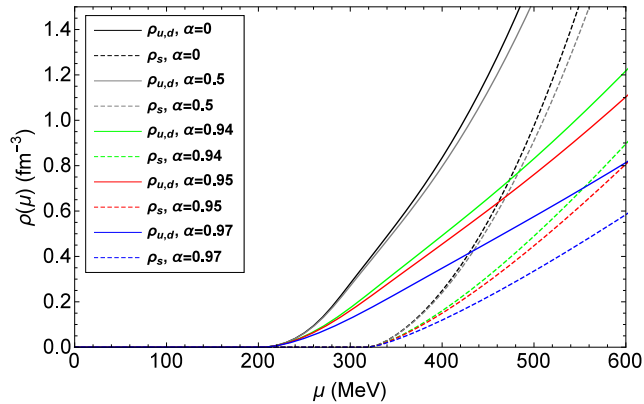}
\caption{Quark number density of $u, d$ and $s$ quarks as a function of $\mu$ at $T=0$ with $\alpha=$0, 0.5, 0.94, 0.95, 0.97, respectively (shown with black, gray, green, red, blue line in correspondence). The densities $\rho_{\rm u,d}$ and $\rho_{\rm s}$ are distinguished by the solid line and dashed line, respectively.}
\label{Fig:qnd}
\end{figure}

By solving Eqs.~(\ref{gapeq}) and (\ref{rechempot}) at zero temperature and finite chemical potential, we can obtain the relation between the quark number density $\langle\psi^+\psi\rangle_{\rm i}$ (also denoted as $\rho_{\rm i}$ in many studies) and its chemical potential $\mu_{\rm i}$, which is shown in Fig.~\ref{Fig:qnd}. In this figure, we can see that for a particular kind of quark, as $\alpha$ changing from 0 to 0.97, the slope of the curve gradually decreases. Specifically, for $\alpha: 0\rightarrow0.5$, the decrease of the slope is very small; but for $\alpha: 0.94\rightarrow0.97$, it seems to be larger. And for the $u$, $d$ quark, the critical chemical potential where the quark number density begins to be nonzero is about 200 MeV; for the s quark, the critical chemical potential is about 320MeV.

\begin{figure}
\includegraphics[width=0.47\textwidth]{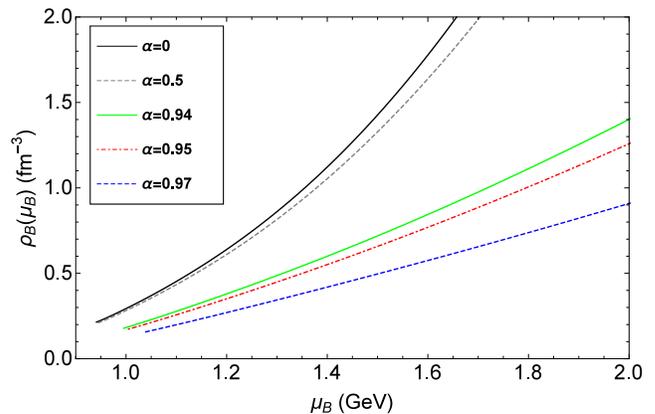}
\caption{Baryon density as a function of baryon chemical potential at $T=0$ with $\alpha=$0, 0.5, 0.94, 0.95, 0.97, respectively. The corresponding curves are black solid, gray dashed, green solid, red dot-dashed and blue dashed, respectively.}
\label{Fig:bd}
\end{figure}

Considering the electro-weak reactions in the quark star, we have to take the chemical equilibrium and the electric charge neutrality into account,
\begin{eqnarray}\label{constrains}
 \mu_{\rm d}&=&\mu_{\rm u}+\mu_{\rm e},\nonumber\\
 \mu_{\rm s}&=&\mu_{\rm u}+\mu_{\rm e},\label{constrains}\\
 \frac{2}{3}\rho_{\rm u}-\frac{1}{3}&\rho_{\rm d}&-\frac{1}{3}\rho_{\rm s}-\rho_{\rm e}=0,\nonumber
\end{eqnarray}
where the electron density at zero temperature reads $\rho_{\rm e}=\mu_{\rm e}^3/(3\pi^2)$. Then the relation of baryon number density $\rho_{\rm B}=(\rho_{\rm u}+\rho_{\rm d}+\rho_{\rm s})/3$ and baryon chemical potential $\mu_{\rm B}=\mu_{\rm u}+\mu_{\rm d}+\mu_{\rm s}$ can be obtained, and the result is shown in Fig.~\ref{Fig:bd}. We can find that the slope of the curves in this figure is also decreasing as $\alpha$ increases.

At zero temperature, the EOS of quark matter can be strictly proved with the functional path integrals~\cite{doi:10.1142/S0217751X08040457,PhysRevD.78.054001},
\begin{equation}\label{EOSofQCD}
P(\mu)=P(\mu=0)+\int_{0}^{\mu}d\mu'\rho(\mu'),
\end{equation}
and the result is model-independent. From Eq.~(\ref{EOSofQCD}) we can find that the pressure of the system can be divided in two parts: one part is 
a density-independent quantity,
i.e., the so-called vacuum pressure, and the other part is density dependent.
Actually, the vacuum pressure $P(\mu=0)$ can not be measured. The only one can be measured is the vacuum pressure difference, and the typical example is the Casimir effect~\cite{PhysRevLett.121.191601,Chernodub:2019nct}. To do this, we need to choose a reference ground. This reference ground state should in principle be a trivial vacuum of the interaction system we are studying~\cite{0954-3899-45-10-105001}.
In the previous studies~\cite{PhysRevD.92.054012,PhysRevD.100.123003,PhysRevD.100.043018}, $P(\mu=0)$ is always taken as a model-dependent phenomenological parameter and associated with -$B$ (vacuum bag constant), just like that in the MIT bag model. However, the value of $B$ should neither be too small nor too large. Because a smaller $B$ corresponds to a stiffer EOS, thus might not meet with constraint on the tidal deformability from GW170817; and a larger $B$ corresponds to a softer EOS, thus might not satisfy the pulsar's mass observation (for example, the PSR J0348+0432 possesses a mass of $2.01\pm0.04 M_{\odot}$~\cite{Antoniadis1233232}, and the recent astronomical observation PSR J0740+6620 provides the most massive neutron star of $2.14_{-0.09}^{+0.10} M_{\odot}$~\cite{Cromartie2019}).

\begin{figure}
\includegraphics[width=0.47\textwidth]{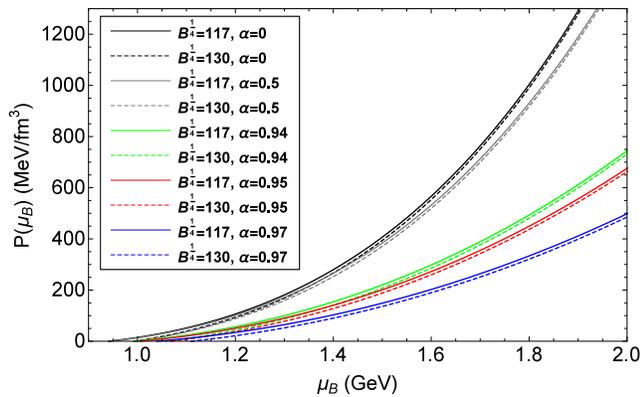}
\caption{The quark EOSs for $\alpha=$0, 0.5, 0.94, 0.95, 0.97 and $B^{\frac{1}{4}}=$117, 130 MeV, respectively. The curves in this figure are plotted in the same type as in Fig.~\ref{Fig:qnd}.}
\label{Fig:EOS}
\end{figure}

In general, the bag constant $B$ has an empirical range of (100 MeV)$^4$-(200 MeV)$^4$~\cite{PhysRevD.46.3211,LU1998443}. And in some recent studies~\cite{PhysRevD.97.083015,PhysRevD.98.083013}, it has been constrained to a narrow range. For example, in Ref.~\cite{PhysRevD.97.083015}, the bag constant is constrained to (134.1, 141.4) MeV based on the study of the quark star with the MIT bag model, and in Ref.~\cite{PhysRevD.98.083013}, it has a parameter space of (166.16, 171.06) MeV in the study of the hybrid star with the NJL model. However, it should be noted that the results above are model-dependent, and the experimental and astronomical observations are still the keys to check whether the value of B we choose is correct at present. In this work, the bag constant is taken as (117 MeV)$^4$, and for comparison, we will also do the calculation for $B=$(130 MeV)$^4$. The result is presented in Fig.~\ref{Fig:EOS}\footnote{To make the captions simple and clear, we omit the units of the parameters captioned in the following figures.}.
In this figure, we can see that the slope of the curves for same $B$ have the same trend as that in Fig.~\ref{Fig:qnd} and Fig.~\ref{Fig:bd}. For the same $\alpha$ but different $B$, a larger $B$ will let the curve move downward along the y-axis.

The energy density and pressure of the system have a relation of~\cite{PhysRevD.86.114028,PhysRevD.51.1989}
\begin{equation}\label{rbedasp}
  \epsilon=-P+\sum_{i}\mu_{\rm i}\rho_{\rm i}.
\end{equation}

Now let us discuss
in the most general sense whether the 2-flavor or the (2+1)-flavor quark matter is more stable.
If we apply Eq.~(\ref{EOSofQCD}) to the 2-flavor and the (2+1)-flavor quark matter system respectively, we can get the following equations,
\begin{equation}\label{NSpressure}
P_{\rm NS}(\mu)=P_{\rm NS}(\mu=0)+\int_{0}^{\mu}{\rm d}\mu'\rho_{\rm NS}(\mu'),
\end{equation}
\begin{equation}\label{Spressure}
P_{\rm S}(\mu)=P_{\rm S}(\mu=0)+\int_{0}^{\mu}{\rm d}\mu'\rho_{\rm S}(\mu'),
\end{equation}
where the subscript ''NS" and ''S" means the non-strange 2-flavor and the strange (2+1)-flavor system, respectively. Subtracting Eq.~(\ref{Spressure}) from Eq.~(\ref{NSpressure}), we can obtain the pressure difference of these two systems,
\begin{eqnarray}
  P_{\rm NS}(\mu)-P_{\rm S}(\mu) =&& [P_{\rm NS}(\mu=0)-P_{\rm S}(\mu=0)]\nonumber\\
   &&+ [\!\!\int_{0}^{\mu}\!\!\!{\rm d}\mu'\rho_{\rm NS}(\mu')-\!\!\int_{0}^{\mu}\!\!\!{\rm d}\mu'\rho_{\rm S}(\mu')].\label{vpdifference}
\end{eqnarray}
It is obvious that at same chemical potential, the system with a higher pressure is more stable than the other one. From Eq.~(\ref{vpdifference}), we can see that the second term to the right side of the equation is density dependent and can be calculated with a certain effective model. However, the first term to the right side of Eq.~(\ref{vpdifference}) is related to the vacuum pressure
difference between the 2-flavor and (2+1)-flavor quark matter,
thus impossible to be calculated from the first principle of QCD,
and this is where we can not judge whether the 2-flavor or the (2+1)-flavor quark matter is more stable.
Therefore, we can not give a theoretically definitive answer to this question at present, and this is also the fundamental reason why Witten's 
strange quark matter hypothesis~\cite{PhysRevD.30.272}
has not yet been proved or falsified. Actually, it might be useful to resort to more and more astronomical observations nowadays to study this question. In some previous studies~\cite{PhysRevD.100.043018,PhysRevD.100.123003,Zhang:2019mqb}, the possibility of the quark star constructed by the 2-flavor quark matter has been discussed, and in this work, we focus on the possibility of the (2+1)-flavor quark matter composing the strange quark star. In addition, we also hope to give an evidence to Witten's
strange quark matter hypothesis~\cite{PhysRevD.30.272}
in this paper with some recent astronomical observations.

In Witten's strange quark matter hypothesis~\cite{PhysRevD.30.272}, the (2+1)-flavor quark system is approbated to be more stable with a lower energy per baryon than the 2-flavor case. Actually, in statistical physics, at the same temperature, a system with a smaller Helmholtz free energy $F$ should be more stable, and the $F$ is just proportional to the energy per baryon, which is demonstrated in the following,
\begin{eqnarray}
  F &=& G_{E}-PV= \sum_{\rm i}(\mu_{\rm i} N_{\rm i}-p_{\rm i}V)\nonumber\\
   &=& V\sum_{\rm i}(\mu_{\rm i} \rho_{\rm i}-p_{\rm i})= V\cdot\epsilon\nonumber\\
   &=& N_{\rm B}\cdot\epsilon/\rho_{\rm B},\,\,\label{HfreeE}
\end{eqnarray}
where $G_{E}$ represents the Gibbs free energy of the system, and $N_{\rm B}=\frac{1}{3}\sum_{\rm i}N_{\rm i}$ is the particle number of baryons. $V$ and $P=\sum_{\rm i}p_{\rm i}$ are the volume and total pressure of the system, respectively. In principle, $P$ and $\epsilon$ should contain the contributions of all constituents of the system, but in the following calculation we ignored the contribution of the electron, as its value is very small compared with the contributions from the deconfined $u,\, d,\, s$ quarks in the quark star.

\begin{figure}
\centering
\subfigure[]{
\centering
\includegraphics[width=0.47\textwidth]{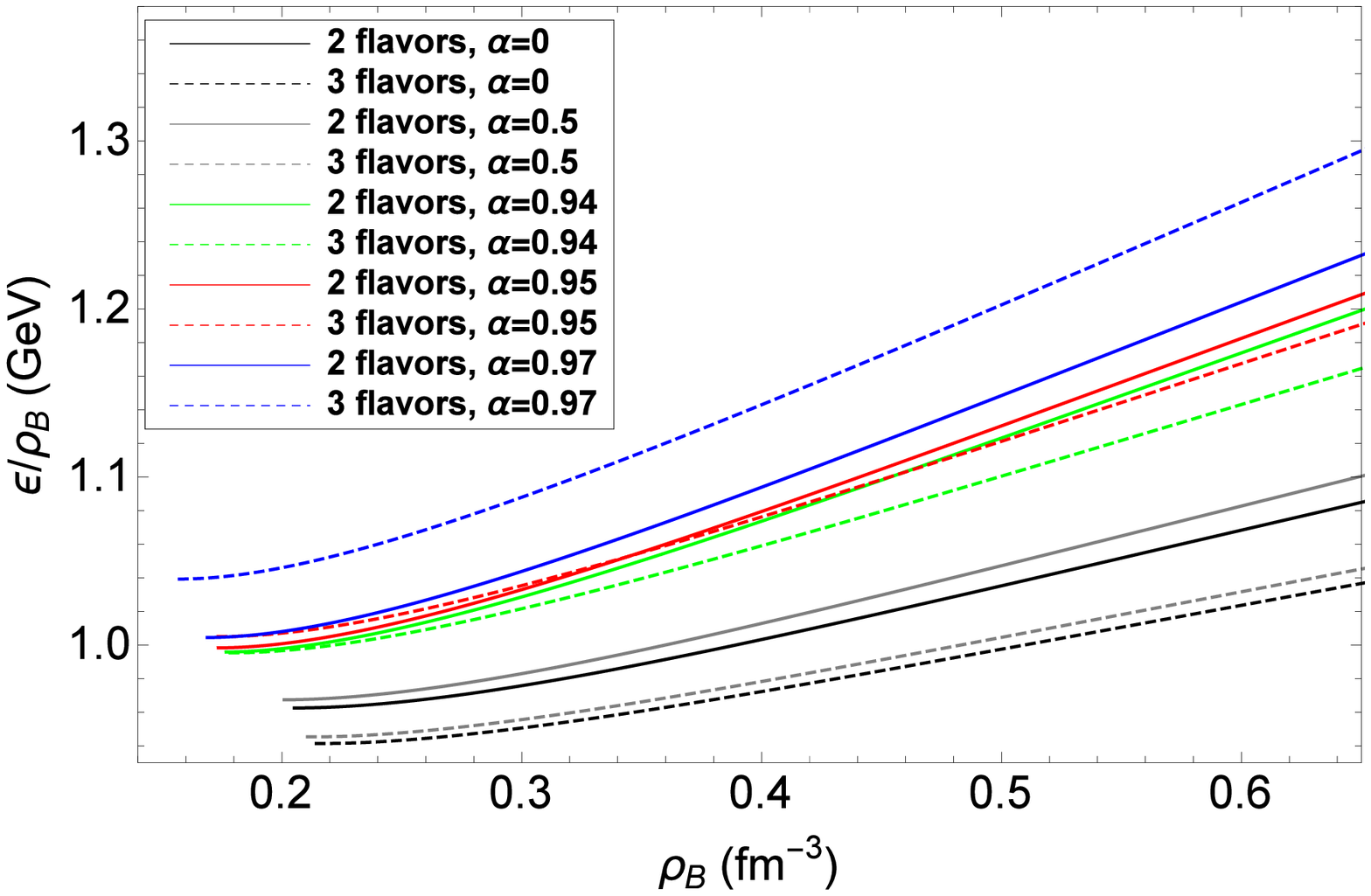}
}
\subfigure[]{
\centering
\includegraphics[width=0.47\textwidth]{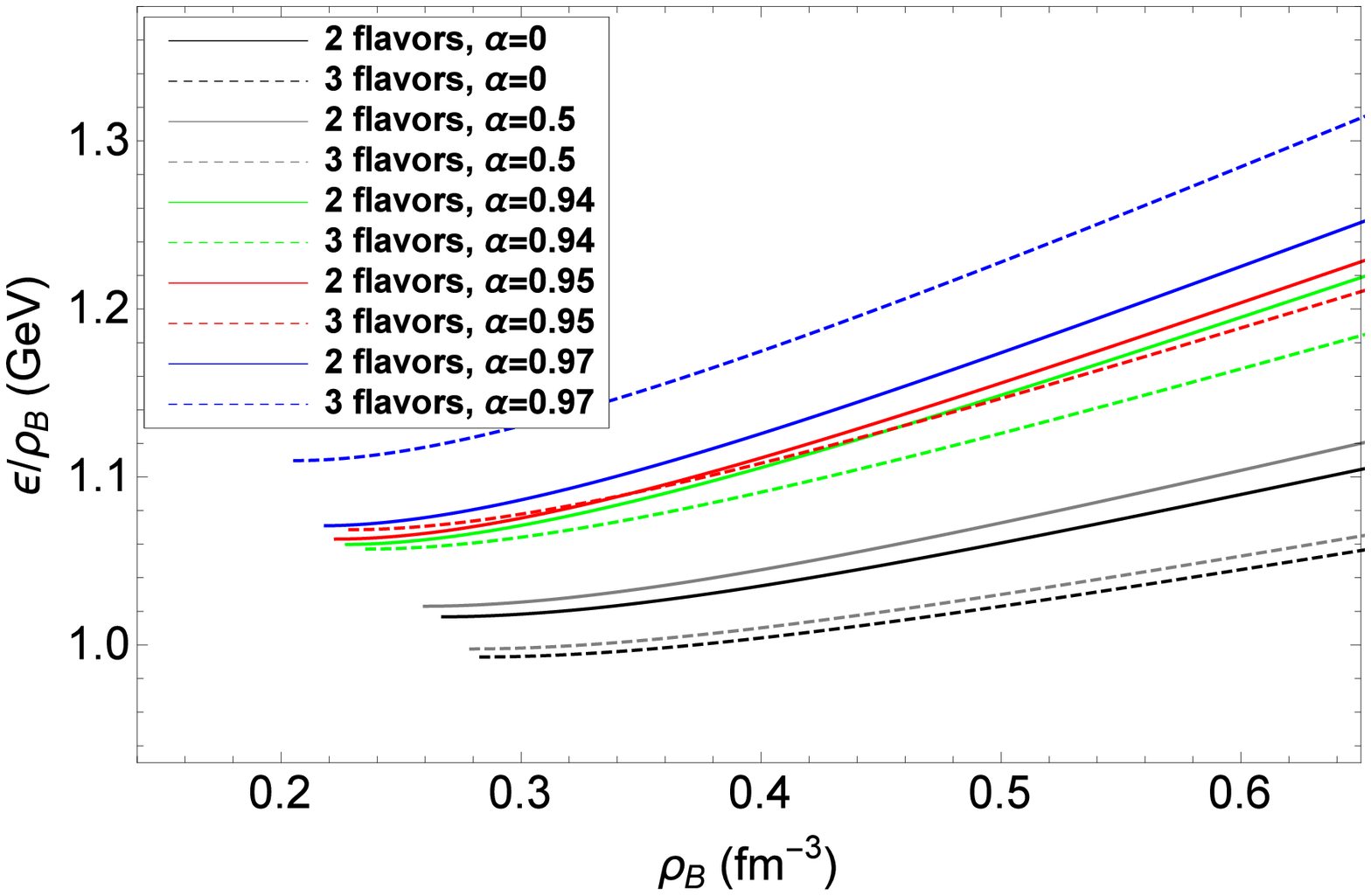}
}
\caption{The comparison of the binding energy of the(2+1)-flavor and 2-flavor system for (a) $B^{\frac{1}{4}}=$117 MeV, and (b) $B^{\frac{1}{4}}=$130 MeV.}
\label{Fig:becomparison}
\end{figure}
It is noted that the $N_{\rm B}$ should be same for the (2+1)-flavor and 2-flavor system if we take the law of conservation of baryon number into account. Thus from Eq.~(\ref{HfreeE}), we can see that a smaller $F$ just corresponds to a smaller energy per baryon $\epsilon/\rho_{\rm B}$ (i.e., the binding energy), and the comparison of the binding energy of these two schemes are shown in Fig.~\ref{Fig:becomparison}\footnote{For sake of consistency, we also use the 2-flavor NJL model with PTR and introduce the Fierz transformation to obtain the binding energy of the 2-flavor system, just as what we do to the 3-flavor system in this work, and for simplicity, the bag constant B is assumed to be the same for these two systems. In addition, the parameter set of the 2-flavor system, ($\Lambda_{\rm UV}$, $G$)=(1330 MeV, $2.028\times10^{-6}$ MeV$^{-2}$), is also fixed under $m_{\rm u}=3.4$ MeV to fit the experimental data ($f_{\pi}$, $M_{\pi}$)=(92, 135) MeV. The specific derivation and calculation process can be referred to Ref.~\cite{PhysRevD.100.123003}.}. It can be found that at the same baryon number density, a larger $B$ will produce a higher binding energy. However, at the same $B$, there are three types of results as $\alpha$ changes: (\romannumeral1), for a small $\alpha$, the binding energy of the (2+1)-flavor system is lower than that of 2-flavor system; (\romannumeral2), for a large $\alpha$, the binding energy of the (2+1)-flavor system is higher than that of 2-flavor system; (\romannumeral3), for a middle $\alpha$ except for the above two situations, the binding energy of the (2+1)-flavor system and that of 2-flavor system intersect, and at the left side of the intersection, that is, at small baryon number densities, the (2+1)-flavor system has a smaller binding energy, but at the right side, the opposite is true. And the results are concluded in Table.~\ref{stability}, actually giving a constraint on the parameter $\alpha$ in this work for the study of the strange quark star. Thus in the following calculation, we will choose some representative values of $\alpha$, i.e., $\alpha=0,\,0.5,\,0.94$ to study the structure of the strange quark star.

\begin{table}
\caption{Comparison of the binding energy of (2+1)-flavor system and 2-flavor system with different $\alpha$, and the system with a lower binding energy is listed in the second and fourth row of the table. The (2+1)-flavor and 2-flavor system are denoted as ``3f" and ``2f", respectively.}\label{stability}
\begin{tabular}{p{1.4cm} p{2cm} p{2.5cm} p{1.6cm}}
\hline\hline
$B^{\frac{1}{4}}$[MeV]&$\,\,\,\,\alpha\leq0.94$&$0.94<\alpha<0.96$&$\,\,\alpha\geq0.96$\\
$\quad$117&$\,\,\,\quad$3f&$\quad\quad$2f$\rightarrow$3f&$\,\,\,\quad$2f\\
\hline
$B^{\frac{1}{4}}$[MeV]&$\,\,\,\,\alpha<0.95$&$0.95\leq\alpha<0.96$&$\,\,\alpha\geq0.96$\\
$\quad$130&$\,\,\,\quad$3f&$\quad\quad$2f$\rightarrow$3f&$\,\,\,\quad$2f\\
\hline\hline
\end{tabular}
\end{table}

\begin{figure}
  \centering
  \includegraphics[width=0.47\textwidth]{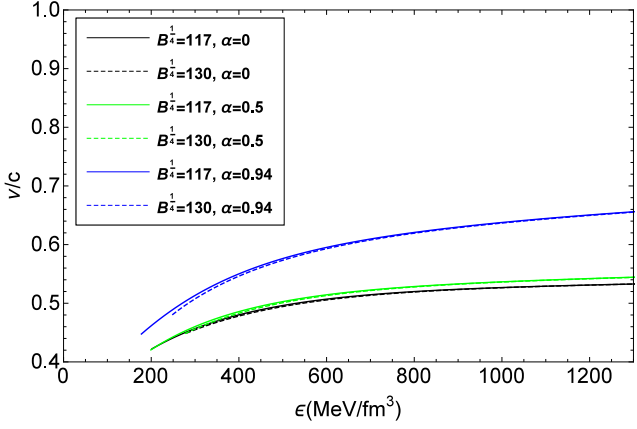}
  \caption{The sound velocities of the six representative strange quark EOSs.}\label{Fig:soundvelocity}
\end{figure}

Now let us investigate the rationality of the strange quark EOS and calculate the sound velocity of it. According to definition, the sound velocity is
\begin{equation}\label{soundvelocity}
 \nu_{\rm s} = \sqrt{\frac{{\rm d}p}{{\rm d}\epsilon}},
\end{equation}
which can reflect the stiffness of the system. Theoretically, a stiffer EOS leads to a larger maximum mass of the compact star, but it might also cause the sound velocity of it exceeds the speed of light, which is unreasonable and should be forbidden. In Fig.~\ref{Fig:soundvelocity}, we show the sound velocities of our six representative strange quark EOSs. In fact, all of them are smaller than 0.7 times speed of light, and a larger $\alpha$ corresponds to a larger sound velocity, demonstrating that the introduction of the Fierz transformation with mean field approximation can make the EOS stiffer compared with the original scheme.

\section{Structure of the strange quark star}\label{two}
\begin{figure}
\includegraphics[width=0.47\textwidth]{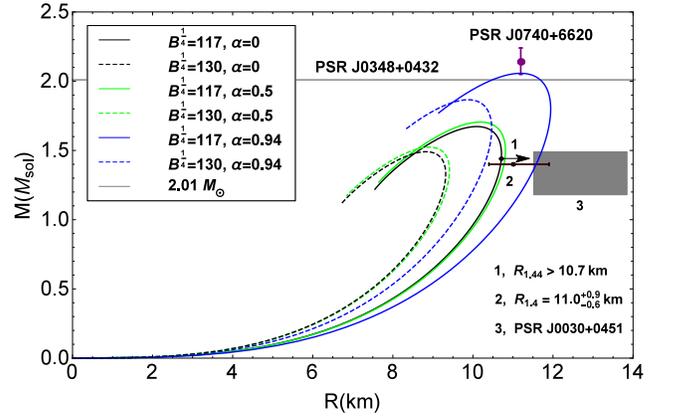}
\caption{The M-R relations of the strange quark star based on the six representative EOSs. Two mass constraints from PSR J0348+0432 and PSR J0740+6620, and three radius constraints in the light of gravitational wave and electro-magnetic observations especially the NICER X-ray timing observations (1, $R_{1.44 M_{\odot}}>10.7$ km~\cite{Bogdanov_2019}; 2, $R_{1.4 M_{\odot}}=11.0^{+0.9}_{-0.6}$ km~\cite{Capano:2019eae}; 3, $M=1.34^{+0.15}_{-0.16}M_{\odot}$, $R=12.71^{+1.14}_{-1.19}$ km from PSR J0030+0451~\cite{Riley_2019}) are also depicted in this figure.}
\label{Fig:mrrelation}
\end{figure}

To get the Mass-radius relation of the quark star, one has to substitute the EOS into the TOV equation and integrate it,
\begin{eqnarray}
  \frac{{\rm d}P(r)}{{\rm d}r} &=& -\frac{G(\epsilon+P)(M+4\pi r^3P)}{r(r-2GM)} \,\, ,\nonumber\\
  \frac{{\rm d}M(r)}{{\rm d}r} &=& 4\pi r^2\epsilon\,\,\, .\label{TOV}
\end{eqnarray}
And the result is shown in Fig.~\ref{Fig:mrrelation}. As a comparison, we also show the mass and radius constraints in this figure based on various observations including the latest pulsar mass~\cite{Cromartie2019,Antoniadis1233232}, X-ray and GW observations~\cite{Bogdanov_2019,Capano:2019eae,Riley_2019}. We can see that only the EOS with $\alpha=0.94$ and $B^{\frac{1}{4}}=117$ MeV meets all five constraints on mass and radius above, producing a maximum mass of 2.057 $M_{\odot}$, and its corresponding radius is about 11.20 km. However, the maximum masses produced by the other EOSs are all smaller than 1.9 $M_{\odot}$. In addition, for the same $B$, a larger $\alpha$ will make the EOS stiffer, generating a larger maximum mass, and for the same $\alpha$, a smaller $B$ also produces a larger maximum mass with a stiffer EOS.

\begin{figure}
\includegraphics[width=0.47\textwidth]{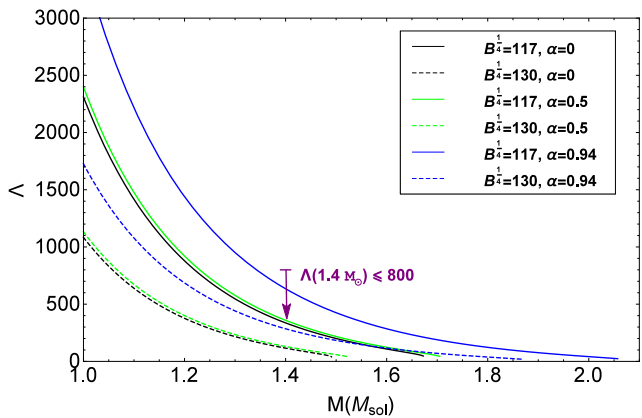}
\caption{The tidal deformability of the strange quark star based on the six representative EOSs, and the constraint on $\Lambda(1.4 M_\odot)$ from GW170817 in the early work~\cite{PhysRevLett.119.161101} is also denoted in this figure.}
\label{Fig:TDresult}
\end{figure}

Then to calculate the tidal deformability, we have to solve the following differential equations along with the solving of the TOV equation, just like what is done in Ref.~\cite{PhysRevD.81.123016},
\begin{eqnarray}
  \frac{dH}{dr} &=& \beta,\nonumber\\
  \frac{d\beta}{dr} &=& 2(1-2\frac{m_r}{r})^{-1}H\{-2\pi[5\epsilon+9p+f(\epsilon+p)]\nonumber\\
  &+&\frac{3}{r^2}+2(1-2\frac{m_r}{r})^{-1}(\frac{m_r}{r^2}+4\pi rp)^2\}\nonumber\\
  &+&\frac{2\beta}{r}(1-2\frac{m_r}{r})^{-1}\{\frac{m_r}{r}+2\pi r^2(\epsilon-p)-1\},\,\label{HbetaEq}
\end{eqnarray}
where H(r) and p are the metric function and pressure, respectively, and $f$ is defined as $d\epsilon/dp$. From this equation we can see that $H$ and its differential equation are also related to the EOS. By defining the quantity $y=R\beta(R)/H(R)-4\pi R^3\epsilon_0/M$, where $\epsilon_0$ represents the energy density at the surface of the quark star, the dimensionless tidal Love number for $l=2$ can be expressed as
\begin{eqnarray}
  &k_2&=\frac{8C^5}{5}(1-2C)^2[2+2C(y-1)-y]\nonumber\\
  &\times&\{2C[6-3y+3C(5y-8)]\nonumber\\
  &+&4C^3[13-11y+C(3y-2)+2C^2(1+y)]\nonumber\\
  &+&3(1-2C)^2[2+2C(y-1)-y]ln(1-2C)\}^{-1},\label{tln}
\end{eqnarray}
where $C=M/R$ refers to the compactness of the quark star. It is noted that the formula of $y$ includes a deduction term $-4\pi R^3\epsilon_0/M$, because the quark matter in the strange quark star is already deconfined, leading to a non-negative pressure at the surface, while the vacuum pressure $P(\mu=0)=-B$ in Eq.~(\ref{EOSofQCD}) is negative, then the quark number density and energy density should be nonzero at the surface. According to Ref.~\cite{PhysRevD.81.123016}, the relation of the tidal deformability $\Lambda$ and the tidal Love number $k_2$ is
\begin{equation}\label{TD}
  \Lambda=\frac{2}{3}k_2R^5.
\end{equation}
Now we can get the result of the tidal deformability of the strange quark star based on the six representative EOSs, which is presented in Fig.~\ref{Fig:TDresult}. We can find that for the quark star mass $M>1$ $M_{\odot}$, as the mass increases, the tidal deformability decreases, and at the same mass, a smaller $B$ or a larger $\alpha$ corresponds to a larger $\Lambda$. And for the EOS producing the maximum mass larger than 1.4 $M_{\odot}$, they all satisfy the constraint for the low-spin priors that $\Lambda(1.4 M_\odot)\leq800$ in the early work~\cite{PhysRevLett.119.161101}.

\begin{figure}
\includegraphics[width=0.47\textwidth]{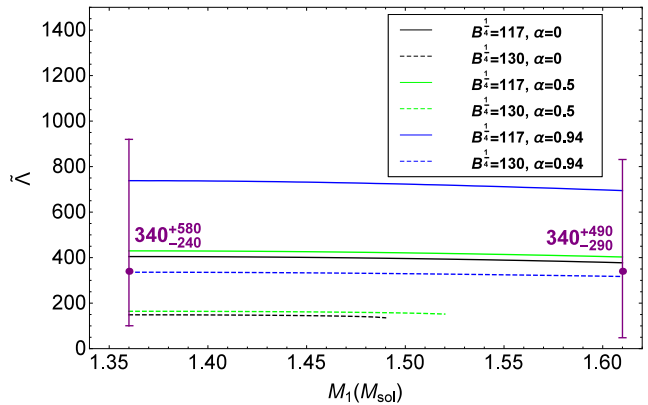}
\caption{The combined tidal deformabilities $\tilde{\Lambda}$ versus the primary star mass $M_1$ of the BNS for the six representative EOSs, and the constraint on $\tilde{\Lambda}$ by GW170817 based on the waveform model TaylorF2, i.e., $\tilde{\Lambda}\sim340^{+580}_{-240}$ for the case of symmetric $90\%$ credible interval and $\tilde{\Lambda}\sim340^{+490}_{-290}$ for the case of highest posterior density (HPD) $90\%$ credible interval.}
\label{Fig:combinedTD}
\end{figure}

Actually, there are some other quantities related to the EOS that are constrained by GW170817, such as the relation of two tidal deformabilities of the BNS, generally presented graphically, and the dimensionless combined tidal deformability $\tilde{\Lambda}$ which is defined as
\begin{equation}\label{combinedTD}
  \tilde{\Lambda}=\frac{16}{13}\frac{(M_1+12M_2)M_1^4\Lambda_1+(M_2+12M_1)M_2^4\Lambda_2}{(M_1+M_2)^5},
\end{equation}
where $M_1$ and $M_2$ are the primary and secondary mass of the BNS, respectively, and their corresponding tidal deformabilities are $\Lambda_1$ and $\Lambda_2$, respectively. For the low-spin priors, in the previous work~\cite{PhysRevLett.119.161101}, the tidal deformability is estimated to be $\tilde{\Lambda}\leq800$ (revised as $\tilde{\Lambda}\leq900$ in Ref.~\cite{PhysRevX.9.011001}), and in the recent work~\cite{PhysRevX.9.011001}, it is restricted to be more accurate, for example, for the waveform model TaylorF2, $\tilde{\Lambda}$ is constrained to be $340^{+580}_{-240}$ for the case of symmetric $90\%$ credible interval and $340^{+490}_{-290}$ for the case of highest posterior density (HPD) $90\%$ credible interval. In Fig.~\ref{Fig:combinedTD}, we show the combined tidal deformabilities of the six representative EOSs, and the constraint from GW170817 based on the waveform model TaylorF2 is also shown in this figure. We can see that the EOSs with ($B^{\frac{1}{4}}$, $\alpha$)=(130, 0.5) and (130, 1) do not satisfy the constraint of HPD, because the maximum masses produced by these two EOSs are smaller than 1.6 $M_{\odot}$ (the maximum mass of the primary star of the BNS), not matching for the requirement of GW170817 in the case of HPD. Among the remaining four EOSs, the one with ($B^{\frac{1}{4}}$, $\alpha$)=(117, 0.94) satisfies not only the mass and radius constraints in Fig.~\ref{Fig:mrrelation}, but also the constraint of combined tidal deformability here. In addition, the $\tilde{\Lambda}$ of the EOSs in this figure only change a little as the mass of the primary star $M_1$ changes, and a smaller $B$ or a larger $\alpha$ corresponds to a larger $\tilde{\Lambda}$.

\begin{figure}
\includegraphics[width=0.47\textwidth]{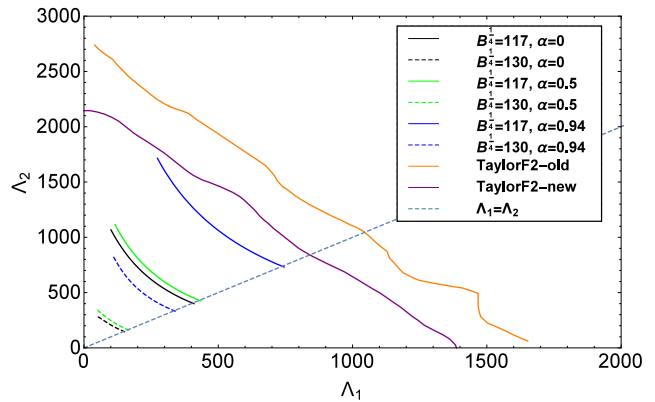}
\caption{The relation of two tidal deformabilities of the BNS, $\Lambda_1-\Lambda_2$ for the representative EOSs. The previous and the recent constraint on $\Lambda_1-\Lambda_2$ via the waveform model TaylorF2 is also denoted in this figure.}
\label{Fig:TDrelation}
\end{figure}

As for the relation of two tidal deformabilities of the BNS, $\Lambda_1-\Lambda_2$, the result is shown in Fig.~\ref{Fig:TDrelation}. In this figure, we can find that even though the constraint on the $\Lambda_1-\Lambda_2$ is improved by the recent study~\cite{PhysRevX.9.011001} compared with the previous one~\cite{PhysRevLett.119.161101}, every representative EOS satisfies the new constraint. Specifically, the $\Lambda_1-\Lambda_2$ relation for the EOS with ($B^{\frac{1}{4}}$, $\alpha$)=(117, 0.94) is just near the edge of the new constraint.

Finally, for sake of completeness, we present the properties of the six strange quark stars based on the representative EOSs in Table.~\ref{sixrepresentative}, including the maximum mass and the corresponding radius, central density as well as the surface density; the radius and tidal deformability of the star with 1.4 $M_{\odot}$ and 1.6 $M_{\odot}$, respectively; the combined tidal deformabilities for the symmetric and HPD case. We can see that among these strange quark EOSs, only the third one with ($B^{\frac{1}{4}}$, $\alpha$)=(117, 0.94) can satisfy the two constraints on the star mass from PSR J0740+6620 and the tidal deformability from GW17017, respectively. Although a larger $\alpha$ can lead to a larger maximum mass, to maintain the (2+1)-flavor quark matter more stable than the 2-flavor case, from Table.~\ref{stability}, the $\alpha$ can not be larger than 0.94 for $B^{\frac{1}{4}}=117$ MeV; on the other hand, a smaller $\alpha$ might let the star mass produced by the EOS fail to meet with the mass constraint of $2.14^{+0.10}_{-0.09}$ $M_{\odot}$. Then how about changing the value of $B^{\frac{1}{4}}$? In Table.~\ref{sixrepresentative} we can see that the increase of $B^{\frac{1}{4}}$ can only reduce the maximum mass, but the decrease of $B^{\frac{1}{4}}$ will make the EOS stiffer, possible to cause the tidal deformability exceeding the constraint from GW170817. For example, in Fig.~\ref{Fig:TDrelation}, the $\Lambda_1-\Lambda_2$ relation curve for ($B^{\frac{1}{4}}$, $\alpha$)=(117, 0.94) is already located near the inner edge of the constraining line of TaylorF2, and we can infer that the replacement to a smaller $B^{\frac{1}{4}}$ is possible to push this curve out. Thus we can conclude that through the introduction of Fierz-transformed Lagrangian to the original NJL Lagrangian, we obtain the suitable strange quark EOS to construct the strange quark star satisfying both the mass and tidal deformability constraint on it, but according to our analysis, the parameter space is still very small, not mentioning the original quark EOS. In a word, these facts suggest that it is reasonable and necessary to introduce Fierz-transformed Lagrangian into the original one.
\begin{widetext}
\begin{center}
\begin{table}
\caption{Some properties of strange quark stars corresponding to the six representative EOSs: maximum gravitational mass $M_{\rm max}$, radius $R_m$, central baryon density $\epsilon_c$, surface baryon density $\epsilon_0$, radius of 1.4 $M_{\odot}$ star $R(1.4)$, tidal deformability of 1.4 $M_{\odot}$ star $\Lambda(1.4)$, radius of 1.6 $M_{\odot}$ star $R(1.6)$, tidal deformability of 1.6 $M_{\odot}$ star $\Lambda(1.6)$, and the combined tidal deformability $\tilde{\Lambda}$ with flat prior (symmetric/HPD).}\label{sixrepresentative}
\begin{tabular}{p{1.0cm} p{1.0cm} p{1.0cm} p{1.0cm} p{1.6cm}p{1.6cm}p{1.2cm}p{1.1cm}p{1.1cm}p{1.1cm}p{2.7cm}}
\hline\hline
$\quad B^{\frac{1}{4}}$&$\quad\alpha$&$M_{\rm max}$&$\,\,R_m$&$\quad\,\,\,\epsilon_c$&$\quad\,\,\,\epsilon_0$&$R(1.4)$&$\Lambda(1.4)$&$R(1.6)$&$\Lambda(1.6)$&$\quad\quad\quad\tilde{\Lambda}$\\
$\,\,[{\rm MeV}]$&$\quad-$&$[M_{\odot}]$&$\,$[km]&$[{\rm MeV/fm^3}]$&$[{\rm MeV/fm^3}]$&$\,\,$[km]&$\quad-$&$\,\,$[km]&$\quad-$&(symmetric/HPD)\\
\hline
\multirow{3}{*}{$\,\,\,\,$117}&$\,\,\,\,\,$0&1.672&10.10&$\quad$1328&$\quad$208&10.69&$\,\,\,$336&10.61&$\,\,\,\,$108&$\quad\quad$405/380\\
                              &$\,\,$0.50&1.705&10.18&$\quad$1315&$\quad$195&10.78&$\,\,\,$359&10.76&$\,\,\,\,$123&$\quad\quad$430/405\\
                              &$\,\,$0.94&2.057&11.20&$\quad$1120&$\quad$182&11.58&$\,\,\,$634&11.83&$\,\,\,\,$285&$\quad\quad$738/698\\
                    \hline
\multirow{3}{*}{$\,\,\,\,$130}&$\,\,\,\,\,$0&1.491&$\,\,$8.83&$\quad$1732&$\quad$286&$\,\,$9.28&$\,\,\,$116&$\,\,\,\,\,-$&$\,\,\,\,\,\,-$&$\,\,\quad\quad$149/$-$\\
                              &$\,\,$0.50&1.523&$\,\,$8.91&$\quad$1706&$\quad$273&$\,\,$9.40&$\,\,\,$130&$\,\,\,\,\,-$&$\,\,\,\,\,\,-$&$\,\,\quad\quad$164/$-$\\
                              &$\,\,$0.94&1.866&$\,\,$9.88&$\quad$1419&$\quad$247&10.29&$\,\,\,$285&10.46&$\,\,\,\,$116&$\quad\quad$336/318\\
\hline\hline
\end{tabular}
\end{table}
\end{center}
\end{widetext}

\section{Summary and discussion}\label{three}
In this paper, to study the EOS and the structure of the strange quark star, we introduce the Fierz-transformed Lagrangian into the original (2+1)-flavor NJL model Lagrangian with the parameter $(1-\alpha)$ and $\alpha$ to combine them linearly. With the mean field approximation and PTR, we fix the parameter set and get the quark number density of the $u$, $d$, $s$ quark. Considering the chemical equilibrium and electric charge neutrality in the star, we get the EOSs with different $\alpha$ and bag constant $B$. To investigate the stableness of the system and make sure that the (2+1)-flavor quark matter is more stable than the 2-flavor case, we compare the binding energies in these two schemes, and find that when $B^{\frac{1}{4}}$=117 MeV, $\alpha\leq$0.94 or $B^{\frac{1}{4}}=130$ MeV, $\alpha<0.95$, the (2+1)-flavor quark system has a smaller binding energy than the 2-flavor one, thus being more stable. Then we calculate six representative strange quark EOSs with $B^{\frac{1}{4}}=117$, 130 MeV and $\alpha=0$, 0.5, 0.94, respectively. And the sound velocities are also calculated to investigate the rationality of them. As a result, none of them exceeds the speed of light and can be adopted for the following calculation.

Then we solve the TOV Eq.~(\ref{TOV}) to get the M-R relation of the strange quark star, and the differential Eq.~(\ref{HbetaEq}) is also solved during this process to obtain the tidal Love number $k_2$ of the star. Via Eq.~(\ref{TD}) and ~(\ref{combinedTD}), the dimensionless tidal deformability $\Lambda$ and combined tidal deformability $\tilde{\Lambda}$ during the BNS merger can also be obtained. Considering the astronomical observation of the neutron star mass and tidal deformability from PSR J0740+6620 and GW170817, respectively, the maximum mass and combined tidal deformability are constrained to be $M_{max}>2.05$ $M_{\odot}$ and ($\tilde{\Lambda}_{symmetric}$, $\tilde{\Lambda}_{HPD}$)$\sim$($340^{+580}_{-240}$, $340^{+490}_{-290}$), respectively. And based on the X-ray observations especially the recent NICER results, the radius of the star is constrained as $R_{1.44 M_{\odot}}>10.7$ km~\cite{Bogdanov_2019}, $R_{1.4 M_{\odot}}=11.0^{+0.9}_{-0.6}$ km~\cite{Capano:2019eae}, and $R=12.71^{+1.14}_{-1.19}$ km with $M=1.34^{+0.15}_{-0.16}M_{\odot}$~\cite{Riley_2019}. Over the six representative EOSs, only the one with parameter ($B^{\frac{1}{4}}$, $\alpha$)=(117, 0.94) can satisfy all the above constraints on mass, radius and tidal deformabilities, reaching to the maximum mass of 2.057 $M_{\odot}$ and the combined tidal deformability is ($\tilde{\Lambda}_{symmetric}$, $\tilde{\Lambda}_{HPD}$)=(770, 724). By analysis, we know that the parameter space is actually very small for our improved EOS model, not mentioning the original one without the introduction of Fierz transformation which can not even yield a star with 2 $M_{\odot}$. Therefore, it is reasonable and necessary to introduce a Fierz-transformed Lagrangian into the original (2+1)-flavor NJL model. On the other hand, our improved EOS model also gives an explanation to the recent neutron star mass and tidal deformability observation from the viewpoint of the strange quark star.

\acknowledgments

This work is supported in part by the National Natural Science Foundation of China (under Grants No. 11475085, No. 11535005, No. 11690030, No. 11473012, No. 11873030 and No. 11675147), the Fundamental Research Funds for the Central Universities (under Grant No. 020414380074), the Strategic Priority Research Program of the Chinese Academy of Sciences "Multi-waveband Gravitational Wave Universe" (Grant No. XDB23040000) and by the National Major state Basic Research and Development of China (Grant No. 2016YFE0129300).

\bibliography{reference}

\begin{thebibliography}{62}%
\makeatletter
\providecommand \@ifxundefined [1]{%
 \@ifx{#1\undefined}
}%
\providecommand \@ifnum [1]{%
 \ifnum #1\expandafter \@firstoftwo
 \else \expandafter \@secondoftwo
 \fi
}%
\providecommand \@ifx [1]{%
 \ifx #1\expandafter \@firstoftwo
 \else \expandafter \@secondoftwo
 \fi
}%
\providecommand \natexlab [1]{#1}%
\providecommand \enquote  [1]{``#1''}%
\providecommand \bibnamefont  [1]{#1}%
\providecommand \bibfnamefont [1]{#1}%
\providecommand \citenamefont [1]{#1}%
\providecommand \href@noop [0]{\@secondoftwo}%
\providecommand \href [0]{\begingroup \@sanitize@url \@href}%
\providecommand \@href[1]{\@@startlink{#1}\@@href}%
\providecommand \@@href[1]{\endgroup#1\@@endlink}%
\providecommand \@sanitize@url [0]{\catcode `\\12\catcode `\$12\catcode
  `\&12\catcode `\#12\catcode `\^12\catcode `\_12\catcode `\%12\relax}%
\providecommand \@@startlink[1]{}%
\providecommand \@@endlink[0]{}%
\providecommand \url  [0]{\begingroup\@sanitize@url \@url }%
\providecommand \@url [1]{\endgroup\@href {#1}{\urlprefix }}%
\providecommand \urlprefix  [0]{URL }%
\providecommand \Eprint [0]{\href }%
\providecommand \doibase [0]{http://dx.doi.org/}%
\providecommand \selectlanguage [0]{\@gobble}%
\providecommand \bibinfo  [0]{\@secondoftwo}%
\providecommand \bibfield  [0]{\@secondoftwo}%
\providecommand \translation [1]{[#1]}%
\providecommand \BibitemOpen [0]{}%
\providecommand \bibitemStop [0]{}%
\providecommand \bibitemNoStop [0]{.\EOS\space}%
\providecommand \EOS [0]{\spacefactor3000\relax}%
\providecommand \BibitemShut  [1]{\csname bibitem#1\endcsname}%
\let\auto@bib@innerbib\@empty
\bibitem [{\citenamefont {Cromartie}\ \emph {et~al.}(2019)\citenamefont
  {Cromartie}, \citenamefont {Fonseca}, \citenamefont {Ransom},\ and\
  \citenamefont {et~al}}]{Cromartie2019}%
  \BibitemOpen
  \bibfield  {author} {\bibinfo {author} {\bibfnamefont {H.~T.}\ \bibnamefont
  {Cromartie}}, \bibinfo {author} {\bibfnamefont {E.}~\bibnamefont {Fonseca}},
  \bibinfo {author} {\bibfnamefont {S.~M.}\ \bibnamefont {Ransom}}, \ and\
  \bibinfo {author} {\bibnamefont {et~al}},\ }\href
  {https://www_nature.xilesou.top/articles/s41550-019-0880-2} {\bibfield
  {journal} {\bibinfo  {journal} {Nat. Astron.}\ }\textbf {\bibinfo {volume}
  {439}} (\bibinfo {year} {2019})}\BibitemShut {NoStop}%
\bibitem [{\citenamefont {Antoniadis}\ \emph {et~al.}(2013)\citenamefont
  {Antoniadis}, \citenamefont {Freire}, \citenamefont {Wex}, \citenamefont
  {Tauris}, \citenamefont {Lynch}, \citenamefont {van Kerkwijk}, \citenamefont
  {Kramer}, \citenamefont {Bassa}, \citenamefont {Dhillon}, \citenamefont
  {Driebe}, \citenamefont {Hessels}, \citenamefont {Kaspi}, \citenamefont
  {Kondratiev}, \citenamefont {Langer}, \citenamefont {Marsh}, \citenamefont
  {McLaughlin}, \citenamefont {Pennucci}, \citenamefont {Ransom}, \citenamefont
  {Stairs}, \citenamefont {van Leeuwen}, \citenamefont {Verbiest},\ and\
  \citenamefont {Whelan}}]{Antoniadis1233232}%
  \BibitemOpen
  \bibfield  {author} {\bibinfo {author} {\bibfnamefont {J.}~\bibnamefont
  {Antoniadis}}, \bibinfo {author} {\bibfnamefont {P.~C.~C.}\ \bibnamefont
  {Freire}}, \bibinfo {author} {\bibfnamefont {N.}~\bibnamefont {Wex}},
  \bibinfo {author} {\bibfnamefont {T.~M.}\ \bibnamefont {Tauris}}, \bibinfo
  {author} {\bibfnamefont {R.~S.}\ \bibnamefont {Lynch}}, \bibinfo {author}
  {\bibfnamefont {M.~H.}\ \bibnamefont {van Kerkwijk}}, \bibinfo {author}
  {\bibfnamefont {M.}~\bibnamefont {Kramer}}, \bibinfo {author} {\bibfnamefont
  {C.}~\bibnamefont {Bassa}}, \bibinfo {author} {\bibfnamefont {V.~S.}\
  \bibnamefont {Dhillon}}, \bibinfo {author} {\bibfnamefont {T.}~\bibnamefont
  {Driebe}}, \bibinfo {author} {\bibfnamefont {J.~W.~T.}\ \bibnamefont
  {Hessels}}, \bibinfo {author} {\bibfnamefont {V.~M.}\ \bibnamefont {Kaspi}},
  \bibinfo {author} {\bibfnamefont {V.~I.}\ \bibnamefont {Kondratiev}},
  \bibinfo {author} {\bibfnamefont {N.}~\bibnamefont {Langer}}, \bibinfo
  {author} {\bibfnamefont {T.~R.}\ \bibnamefont {Marsh}}, \bibinfo {author}
  {\bibfnamefont {M.~A.}\ \bibnamefont {McLaughlin}}, \bibinfo {author}
  {\bibfnamefont {T.~T.}\ \bibnamefont {Pennucci}}, \bibinfo {author}
  {\bibfnamefont {S.~M.}\ \bibnamefont {Ransom}}, \bibinfo {author}
  {\bibfnamefont {I.~H.}\ \bibnamefont {Stairs}}, \bibinfo {author}
  {\bibfnamefont {J.}~\bibnamefont {van Leeuwen}}, \bibinfo {author}
  {\bibfnamefont {J.~P.~W.}\ \bibnamefont {Verbiest}}, \ and\ \bibinfo {author}
  {\bibfnamefont {D.~G.}\ \bibnamefont {Whelan}},\ }\href
  {https://science.sciencemag.org/cgi/doi/10.1126/science.1233232} {\bibfield
  {journal} {\bibinfo  {journal} {Science}\ }\textbf {\bibinfo {volume}
  {340}},\ \bibinfo {pages} {1233232} (\bibinfo {year} {2013})}\BibitemShut
  {NoStop}%
\bibitem [{\citenamefont {Bogdanov}\ \emph {et~al.}(2019)\citenamefont
  {Bogdanov}, \citenamefont {Guillot}, \citenamefont {Ray}, \citenamefont
  {Wolff}, \citenamefont {Chakrabarty}, \citenamefont {Ho}, \citenamefont
  {Kerr}, \citenamefont {Lamb}, \citenamefont {Lommen}, \citenamefont {Ludlam},
  \citenamefont {Milburn}, \citenamefont {Montano}, \citenamefont {Miller},
  \citenamefont {Baubock}, \citenamefont {Ozel}, \citenamefont {Psaltis},
  \citenamefont {Remillard}, \citenamefont {Riley}, \citenamefont {Steiner},
  \citenamefont {Strohmayer}, \citenamefont {Watts}, \citenamefont {Wood},
  \citenamefont {Zeldes}, \citenamefont {Enoto}, \citenamefont {Okajima},
  \citenamefont {Kellogg}, \citenamefont {Baker}, \citenamefont {Markwardt},
  \citenamefont {Arzoumanian},\ and\ \citenamefont {Gendreau}}]{Bogdanov_2019}%
  \BibitemOpen
  \bibfield  {author} {\bibinfo {author} {\bibfnamefont {S.}~\bibnamefont
  {Bogdanov}}, \bibinfo {author} {\bibfnamefont {S.}~\bibnamefont {Guillot}},
  \bibinfo {author} {\bibfnamefont {P.~S.}\ \bibnamefont {Ray}}, \bibinfo
  {author} {\bibfnamefont {M.~T.}\ \bibnamefont {Wolff}}, \bibinfo {author}
  {\bibfnamefont {D.}~\bibnamefont {Chakrabarty}}, \bibinfo {author}
  {\bibfnamefont {W.~C.~G.}\ \bibnamefont {Ho}}, \bibinfo {author}
  {\bibfnamefont {M.}~\bibnamefont {Kerr}}, \bibinfo {author} {\bibfnamefont
  {F.~K.}\ \bibnamefont {Lamb}}, \bibinfo {author} {\bibfnamefont
  {A.}~\bibnamefont {Lommen}}, \bibinfo {author} {\bibfnamefont {R.~M.}\
  \bibnamefont {Ludlam}}, \bibinfo {author} {\bibfnamefont {R.}~\bibnamefont
  {Milburn}}, \bibinfo {author} {\bibfnamefont {S.}~\bibnamefont {Montano}},
  \bibinfo {author} {\bibfnamefont {M.~C.}\ \bibnamefont {Miller}}, \bibinfo
  {author} {\bibfnamefont {M.}~\bibnamefont {Baubock}}, \bibinfo {author}
  {\bibfnamefont {F.}~\bibnamefont {Ozel}}, \bibinfo {author} {\bibfnamefont
  {D.}~\bibnamefont {Psaltis}}, \bibinfo {author} {\bibfnamefont {R.~A.}\
  \bibnamefont {Remillard}}, \bibinfo {author} {\bibfnamefont {T.~E.}\
  \bibnamefont {Riley}}, \bibinfo {author} {\bibfnamefont {J.~F.}\ \bibnamefont
  {Steiner}}, \bibinfo {author} {\bibfnamefont {T.~E.}\ \bibnamefont
  {Strohmayer}}, \bibinfo {author} {\bibfnamefont {A.~L.}\ \bibnamefont
  {Watts}}, \bibinfo {author} {\bibfnamefont {K.~S.}\ \bibnamefont {Wood}},
  \bibinfo {author} {\bibfnamefont {J.}~\bibnamefont {Zeldes}}, \bibinfo
  {author} {\bibfnamefont {T.}~\bibnamefont {Enoto}}, \bibinfo {author}
  {\bibfnamefont {T.}~\bibnamefont {Okajima}}, \bibinfo {author} {\bibfnamefont
  {J.~W.}\ \bibnamefont {Kellogg}}, \bibinfo {author} {\bibfnamefont
  {C.}~\bibnamefont {Baker}}, \bibinfo {author} {\bibfnamefont {C.~B.}\
  \bibnamefont {Markwardt}}, \bibinfo {author} {\bibfnamefont {Z.}~\bibnamefont
  {Arzoumanian}}, \ and\ \bibinfo {author} {\bibfnamefont {K.~C.}\ \bibnamefont
  {Gendreau}},\ }\href {\doibase 10.3847/2041-8213/ab53eb} {\bibfield
  {journal} {\bibinfo  {journal} {Astrophys. J}\ }\textbf {\bibinfo {volume}
  {887}},\ \bibinfo {pages} {L25} (\bibinfo {year} {2019})}\BibitemShut
  {NoStop}%
\bibitem [{\citenamefont {Capano}\ \emph {et~al.}(2019)\citenamefont {Capano},
  \citenamefont {Tews}, \citenamefont {Brown}, \citenamefont {Margalit},
  \citenamefont {De}, \citenamefont {Kumar}, \citenamefont {Brown},
  \citenamefont {Krishnan},\ and\ \citenamefont {Reddy}}]{Capano:2019eae}%
  \BibitemOpen
  \bibfield  {author} {\bibinfo {author} {\bibfnamefont {C.~D.}\ \bibnamefont
  {Capano}}, \bibinfo {author} {\bibfnamefont {I.}~\bibnamefont {Tews}},
  \bibinfo {author} {\bibfnamefont {S.~M.}\ \bibnamefont {Brown}}, \bibinfo
  {author} {\bibfnamefont {B.}~\bibnamefont {Margalit}}, \bibinfo {author}
  {\bibfnamefont {S.}~\bibnamefont {De}}, \bibinfo {author} {\bibfnamefont
  {S.}~\bibnamefont {Kumar}}, \bibinfo {author} {\bibfnamefont {D.~A.}\
  \bibnamefont {Brown}}, \bibinfo {author} {\bibfnamefont {B.}~\bibnamefont
  {Krishnan}}, \ and\ \bibinfo {author} {\bibfnamefont {S.}~\bibnamefont
  {Reddy}},\ }\href@noop {} {\  (\bibinfo {year} {2019})},\ \Eprint
  {http://arxiv.org/abs/1908.10352} {arXiv:1908.10352 [astro-ph.HE]}
  \BibitemShut {NoStop}%
\bibitem [{\citenamefont {Riley}\ \emph {et~al.}(2019)\citenamefont {Riley},
  \citenamefont {Watts}, \citenamefont {Bogdanov}, \citenamefont {Ray},
  \citenamefont {Ludlam}, \citenamefont {Guillot}, \citenamefont {Arzoumanian},
  \citenamefont {Baker}, \citenamefont {Bilous}, \citenamefont {Chakrabarty},
  \citenamefont {Gendreau}, \citenamefont {Harding}, \citenamefont {Ho},
  \citenamefont {Lattimer}, \citenamefont {Morsink},\ and\ \citenamefont
  {Strohmayer}}]{Riley_2019}%
  \BibitemOpen
  \bibfield  {author} {\bibinfo {author} {\bibfnamefont {T.~E.}\ \bibnamefont
  {Riley}}, \bibinfo {author} {\bibfnamefont {A.~L.}\ \bibnamefont {Watts}},
  \bibinfo {author} {\bibfnamefont {S.}~\bibnamefont {Bogdanov}}, \bibinfo
  {author} {\bibfnamefont {P.~S.}\ \bibnamefont {Ray}}, \bibinfo {author}
  {\bibfnamefont {R.~M.}\ \bibnamefont {Ludlam}}, \bibinfo {author}
  {\bibfnamefont {S.}~\bibnamefont {Guillot}}, \bibinfo {author} {\bibfnamefont
  {Z.}~\bibnamefont {Arzoumanian}}, \bibinfo {author} {\bibfnamefont {C.~L.}\
  \bibnamefont {Baker}}, \bibinfo {author} {\bibfnamefont {A.~V.}\ \bibnamefont
  {Bilous}}, \bibinfo {author} {\bibfnamefont {D.}~\bibnamefont {Chakrabarty}},
  \bibinfo {author} {\bibfnamefont {K.~C.}\ \bibnamefont {Gendreau}}, \bibinfo
  {author} {\bibfnamefont {A.~K.}\ \bibnamefont {Harding}}, \bibinfo {author}
  {\bibfnamefont {W.~C.~G.}\ \bibnamefont {Ho}}, \bibinfo {author}
  {\bibfnamefont {J.~M.}\ \bibnamefont {Lattimer}}, \bibinfo {author}
  {\bibfnamefont {S.~M.}\ \bibnamefont {Morsink}}, \ and\ \bibinfo {author}
  {\bibfnamefont {T.~E.}\ \bibnamefont {Strohmayer}},\ }\href {\doibase
  10.3847/2041-8213/ab481c} {\bibfield  {journal} {\bibinfo  {journal}
  {Astrophys. J}\ }\textbf {\bibinfo {volume} {887}},\ \bibinfo {pages} {L21}
  (\bibinfo {year} {2019})}\BibitemShut {NoStop}%
\bibitem [{\citenamefont {Abbott}\ and\ \citenamefont
  {et~al.}(2017)}]{PhysRevLett.119.161101}%
  \BibitemOpen
  \bibfield  {author} {\bibinfo {author} {\bibfnamefont {B.~P.}\ \bibnamefont
  {Abbott}}\ and\ \bibinfo {author} {\bibnamefont {et~al.}} (\bibinfo
  {collaboration} {LIGO Scientific Collaboration and Virgo Collaboration}),\
  }\href {\doibase 10.1103/PhysRevLett.119.161101} {\bibfield  {journal}
  {\bibinfo  {journal} {Phys. Rev. Lett.}\ }\textbf {\bibinfo {volume} {119}},\
  \bibinfo {pages} {161101} (\bibinfo {year} {2017})}\BibitemShut {NoStop}%
\bibitem [{\citenamefont {Abbott}\ and\ \citenamefont
  {et~al.}(2019)}]{PhysRevX.9.011001}%
  \BibitemOpen
  \bibfield  {author} {\bibinfo {author} {\bibfnamefont {B.~P.}\ \bibnamefont
  {Abbott}}\ and\ \bibinfo {author} {\bibnamefont {et~al.}} (\bibinfo
  {collaboration} {LIGO Scientific Collaboration and Virgo Collaboration}),\
  }\href {\doibase 10.1103/PhysRevX.9.011001} {\bibfield  {journal} {\bibinfo
  {journal} {Phys. Rev. X}\ }\textbf {\bibinfo {volume} {9}},\ \bibinfo {pages}
  {011001} (\bibinfo {year} {2019})}\BibitemShut {NoStop}%
\bibitem [{\citenamefont {Witten}(1984)}]{PhysRevD.30.272}%
  \BibitemOpen
  \bibfield  {author} {\bibinfo {author} {\bibfnamefont {E.}~\bibnamefont
  {Witten}},\ }\href {\doibase 10.1103/PhysRevD.30.272} {\bibfield  {journal}
  {\bibinfo  {journal} {Phys. Rev. D}\ }\textbf {\bibinfo {volume} {30}},\
  \bibinfo {pages} {272} (\bibinfo {year} {1984})}\BibitemShut {NoStop}%
\bibitem [{\citenamefont {Dexheimer}\ \emph {et~al.}(2013)\citenamefont
  {Dexheimer}, \citenamefont {Torres},\ and\ \citenamefont
  {Menezes}}]{Dexheimer2013}%
  \BibitemOpen
  \bibfield  {author} {\bibinfo {author} {\bibfnamefont {V.}~\bibnamefont
  {Dexheimer}}, \bibinfo {author} {\bibfnamefont {J.~R.}\ \bibnamefont
  {Torres}}, \ and\ \bibinfo {author} {\bibfnamefont {D.~P.}\ \bibnamefont
  {Menezes}},\ }\href {\doibase 10.1140/epjc/s10052-013-2569-5} {\bibfield
  {journal} {\bibinfo  {journal} {The European Physical Journal C}\ }\textbf
  {\bibinfo {volume} {73}},\ \bibinfo {pages} {2569} (\bibinfo {year}
  {2013})}\BibitemShut {NoStop}%
\bibitem [{\citenamefont
  {Terazawa}(1989{\natexlab{a}})}]{doi:10.1143/JPSJ.58.3555}%
  \BibitemOpen
  \bibfield  {author} {\bibinfo {author} {\bibfnamefont {H.}~\bibnamefont
  {Terazawa}},\ }\href {\doibase 10.1143/JPSJ.58.3555} {\bibfield  {journal}
  {\bibinfo  {journal} {Journal of the Physical Society of Japan}\ }\textbf
  {\bibinfo {volume} {58}},\ \bibinfo {pages} {3555} (\bibinfo {year}
  {1989}{\natexlab{a}})}\BibitemShut {NoStop}%
\bibitem [{\citenamefont
  {Terazawa}(1989{\natexlab{b}})}]{doi:10.1143/JPSJ.58.4388}%
  \BibitemOpen
  \bibfield  {author} {\bibinfo {author} {\bibfnamefont {H.}~\bibnamefont
  {Terazawa}},\ }\href {\doibase 10.1143/JPSJ.58.4388} {\bibfield  {journal}
  {\bibinfo  {journal} {Journal of the Physical Society of Japan}\ }\textbf
  {\bibinfo {volume} {58}},\ \bibinfo {pages} {4388} (\bibinfo {year}
  {1989}{\natexlab{b}})}\BibitemShut {NoStop}%
\bibitem [{\citenamefont {Holdom}\ \emph {et~al.}(2018)\citenamefont {Holdom},
  \citenamefont {Ren},\ and\ \citenamefont {Zhang}}]{PhysRevLett.120.222001}%
  \BibitemOpen
  \bibfield  {author} {\bibinfo {author} {\bibfnamefont {B.}~\bibnamefont
  {Holdom}}, \bibinfo {author} {\bibfnamefont {J.}~\bibnamefont {Ren}}, \ and\
  \bibinfo {author} {\bibfnamefont {C.}~\bibnamefont {Zhang}},\ }\href
  {\doibase 10.1103/PhysRevLett.120.222001} {\bibfield  {journal} {\bibinfo
  {journal} {Phys. Rev. Lett.}\ }\textbf {\bibinfo {volume} {120}},\ \bibinfo
  {pages} {222001} (\bibinfo {year} {2018})}\BibitemShut {NoStop}%
\bibitem [{\citenamefont {Zhao}\ \emph {et~al.}(2019)\citenamefont {Zhao},
  \citenamefont {Zheng}, \citenamefont {Wang}, \citenamefont {Li},
  \citenamefont {Yan}, \citenamefont {Huang},\ and\ \citenamefont
  {Zong}}]{PhysRevD.100.043018}%
  \BibitemOpen
  \bibfield  {author} {\bibinfo {author} {\bibfnamefont {T.}~\bibnamefont
  {Zhao}}, \bibinfo {author} {\bibfnamefont {W.}~\bibnamefont {Zheng}},
  \bibinfo {author} {\bibfnamefont {F.}~\bibnamefont {Wang}}, \bibinfo {author}
  {\bibfnamefont {C.-M.}\ \bibnamefont {Li}}, \bibinfo {author} {\bibfnamefont
  {Y.}~\bibnamefont {Yan}}, \bibinfo {author} {\bibfnamefont {Y.-F.}\
  \bibnamefont {Huang}}, \ and\ \bibinfo {author} {\bibfnamefont {H.-S.}\
  \bibnamefont {Zong}},\ }\href {\doibase 10.1103/PhysRevD.100.043018}
  {\bibfield  {journal} {\bibinfo  {journal} {Phys. Rev. D}\ }\textbf {\bibinfo
  {volume} {100}},\ \bibinfo {pages} {043018} (\bibinfo {year}
  {2019})}\BibitemShut {NoStop}%
\bibitem [{\citenamefont {Wang}\ \emph
  {et~al.}(2019{\natexlab{a}})\citenamefont {Wang}, \citenamefont {Shi},\ and\
  \citenamefont {Zong}}]{PhysRevD.100.123003}%
  \BibitemOpen
  \bibfield  {author} {\bibinfo {author} {\bibfnamefont {Q.}~\bibnamefont
  {Wang}}, \bibinfo {author} {\bibfnamefont {C.}~\bibnamefont {Shi}}, \ and\
  \bibinfo {author} {\bibfnamefont {H.-S.}\ \bibnamefont {Zong}},\ }\href
  {\doibase 10.1103/PhysRevD.100.123003} {\bibfield  {journal} {\bibinfo
  {journal} {Phys. Rev. D}\ }\textbf {\bibinfo {volume} {100}},\ \bibinfo
  {pages} {123003} (\bibinfo {year} {2019}{\natexlab{a}})}\BibitemShut
  {NoStop}%
\bibitem [{\citenamefont {Zhang}(2019)}]{Zhang:2019mqb}%
  \BibitemOpen
  \bibfield  {author} {\bibinfo {author} {\bibfnamefont {C.}~\bibnamefont
  {Zhang}},\ }\href@noop {} {\  (\bibinfo {year} {2019})},\ \Eprint
  {http://arxiv.org/abs/1908.10355} {arXiv:1908.10355 [astro-ph.HE]}
  \BibitemShut {NoStop}%
\bibitem [{\citenamefont {Roberts}\ and\ \citenamefont
  {Williams}(1994)}]{ROBERTS1994477}%
  \BibitemOpen
  \bibfield  {author} {\bibinfo {author} {\bibfnamefont {C.~D.}\ \bibnamefont
  {Roberts}}\ and\ \bibinfo {author} {\bibfnamefont {A.~G.}\ \bibnamefont
  {Williams}},\ }\href {\doibase
  http://dx.doi.org/10.1016/0146-6410(94)90049-3} {\bibfield  {journal}
  {\bibinfo  {journal} {Prog. Part. Nucl. Phys.}\ }\textbf {\bibinfo {volume}
  {33}},\ \bibinfo {pages} {477 } (\bibinfo {year} {1994})}\BibitemShut
  {NoStop}%
\bibitem [{\citenamefont {Roberts}\ and\ \citenamefont
  {Schmidt}(2000)}]{Roberts2000S1}%
  \BibitemOpen
  \bibfield  {author} {\bibinfo {author} {\bibfnamefont {C.}~\bibnamefont
  {Roberts}}\ and\ \bibinfo {author} {\bibfnamefont {S.}~\bibnamefont
  {Schmidt}},\ }\href {\doibase
  http://dx.doi.org/10.1016/S0146-6410(00)90011-5} {\bibfield  {journal}
  {\bibinfo  {journal} {Prog. Part. Nucl. Phys.}\ }\textbf {\bibinfo {volume}
  {45, Supplement 1}},\ \bibinfo {pages} {S1 } (\bibinfo {year}
  {2000})}\BibitemShut {NoStop}%
\bibitem [{\citenamefont {Maris}\ and\ \citenamefont
  {Roberts}(2003)}]{doi:10.1142/S0218301303001326}%
  \BibitemOpen
  \bibfield  {author} {\bibinfo {author} {\bibfnamefont {P.}~\bibnamefont
  {Maris}}\ and\ \bibinfo {author} {\bibfnamefont {C.~D.}\ \bibnamefont
  {Roberts}},\ }\href {\doibase 10.1142/S0218301303001326} {\bibfield
  {journal} {\bibinfo  {journal} {Int. J. Mod. Phys. E}\ }\textbf {\bibinfo
  {volume} {12}},\ \bibinfo {pages} {297} (\bibinfo {year} {2003})}\BibitemShut
  {NoStop}%
\bibitem [{\citenamefont {Cl{\"o}et}\ and\ \citenamefont
  {Roberts}(2014)}]{Cloet20141}%
  \BibitemOpen
  \bibfield  {author} {\bibinfo {author} {\bibfnamefont {I.~C.}\ \bibnamefont
  {Cl{\"o}et}}\ and\ \bibinfo {author} {\bibfnamefont {C.~D.}\ \bibnamefont
  {Roberts}},\ }\href {\doibase http://dx.doi.org/10.1016/j.ppnp.2014.02.001}
  {\bibfield  {journal} {\bibinfo  {journal} {Prog. Part. Nucl. Phys.}\
  }\textbf {\bibinfo {volume} {77}},\ \bibinfo {pages} {1 } (\bibinfo {year}
  {2014})}\BibitemShut {NoStop}%
\bibitem [{\citenamefont {Zhao}\ \emph {et~al.}(2014)\citenamefont {Zhao},
  \citenamefont {Cui}, \citenamefont {Jiang},\ and\ \citenamefont
  {Zong}}]{PhysRevD.90.114031}%
  \BibitemOpen
  \bibfield  {author} {\bibinfo {author} {\bibfnamefont {A.-M.}\ \bibnamefont
  {Zhao}}, \bibinfo {author} {\bibfnamefont {Z.-F.}\ \bibnamefont {Cui}},
  \bibinfo {author} {\bibfnamefont {Y.}~\bibnamefont {Jiang}}, \ and\ \bibinfo
  {author} {\bibfnamefont {H.-S.}\ \bibnamefont {Zong}},\ }\href {\doibase
  10.1103/PhysRevD.90.114031} {\bibfield  {journal} {\bibinfo  {journal} {Phys.
  Rev. D}\ }\textbf {\bibinfo {volume} {90}},\ \bibinfo {pages} {114031}
  (\bibinfo {year} {2014})}\BibitemShut {NoStop}%
\bibitem [{\citenamefont {Wang}\ \emph {et~al.}(2015)\citenamefont {Wang},
  \citenamefont {Wang}, \citenamefont {Cui},\ and\ \citenamefont
  {Zong}}]{PhysRevD.91.034017}%
  \BibitemOpen
  \bibfield  {author} {\bibinfo {author} {\bibfnamefont {B.}~\bibnamefont
  {Wang}}, \bibinfo {author} {\bibfnamefont {Y.-L.}\ \bibnamefont {Wang}},
  \bibinfo {author} {\bibfnamefont {Z.-F.}\ \bibnamefont {Cui}}, \ and\
  \bibinfo {author} {\bibfnamefont {H.-S.}\ \bibnamefont {Zong}},\ }\href
  {\doibase 10.1103/PhysRevD.91.034017} {\bibfield  {journal} {\bibinfo
  {journal} {Phys. Rev. D}\ }\textbf {\bibinfo {volume} {91}},\ \bibinfo
  {pages} {034017} (\bibinfo {year} {2015})}\BibitemShut {NoStop}%
\bibitem [{\citenamefont {Xu}\ \emph {et~al.}(2015)\citenamefont {Xu},
  \citenamefont {Cui}, \citenamefont {Wang}, \citenamefont {Shi}, \citenamefont
  {Yang},\ and\ \citenamefont {Zong}}]{PhysRevD.91.056003}%
  \BibitemOpen
  \bibfield  {author} {\bibinfo {author} {\bibfnamefont {S.-S.}\ \bibnamefont
  {Xu}}, \bibinfo {author} {\bibfnamefont {Z.-F.}\ \bibnamefont {Cui}},
  \bibinfo {author} {\bibfnamefont {B.}~\bibnamefont {Wang}}, \bibinfo {author}
  {\bibfnamefont {Y.-M.}\ \bibnamefont {Shi}}, \bibinfo {author} {\bibfnamefont
  {Y.-C.}\ \bibnamefont {Yang}}, \ and\ \bibinfo {author} {\bibfnamefont
  {H.-S.}\ \bibnamefont {Zong}},\ }\href {\doibase 10.1103/PhysRevD.91.056003}
  {\bibfield  {journal} {\bibinfo  {journal} {Phys. Rev. D}\ }\textbf {\bibinfo
  {volume} {91}},\ \bibinfo {pages} {056003} (\bibinfo {year}
  {2015})}\BibitemShut {NoStop}%
\bibitem [{\citenamefont {Pisarski}(1984)}]{PhysRevD.29.2423}%
  \BibitemOpen
  \bibfield  {author} {\bibinfo {author} {\bibfnamefont {R.~D.}\ \bibnamefont
  {Pisarski}},\ }\href {\doibase 10.1103/PhysRevD.29.2423} {\bibfield
  {journal} {\bibinfo  {journal} {Phys. Rev. D}\ }\textbf {\bibinfo {volume}
  {29}},\ \bibinfo {pages} {2423} (\bibinfo {year} {1984})}\BibitemShut
  {NoStop}%
\bibitem [{\citenamefont {Yin}\ \emph {et~al.}(2014)\citenamefont {Yin},
  \citenamefont {Shi}, \citenamefont {Cui}, \citenamefont {Feng},\ and\
  \citenamefont {Zong}}]{PhysRevD.90.036007}%
  \BibitemOpen
  \bibfield  {author} {\bibinfo {author} {\bibfnamefont {P.-L.}\ \bibnamefont
  {Yin}}, \bibinfo {author} {\bibfnamefont {Y.-M.}\ \bibnamefont {Shi}},
  \bibinfo {author} {\bibfnamefont {Z.-F.}\ \bibnamefont {Cui}}, \bibinfo
  {author} {\bibfnamefont {H.-T.}\ \bibnamefont {Feng}}, \ and\ \bibinfo
  {author} {\bibfnamefont {H.-S.}\ \bibnamefont {Zong}},\ }\href {\doibase
  10.1103/PhysRevD.90.036007} {\bibfield  {journal} {\bibinfo  {journal} {Phys.
  Rev. D}\ }\textbf {\bibinfo {volume} {90}},\ \bibinfo {pages} {036007}
  (\bibinfo {year} {2014})}\BibitemShut {NoStop}%
\bibitem [{\citenamefont {Li}\ \emph {et~al.}(2014)\citenamefont {Li},
  \citenamefont {Hou}, \citenamefont {Cui}, \citenamefont {Feng}, \citenamefont
  {Jiang},\ and\ \citenamefont {Zong}}]{PhysRevD.90.073013}%
  \BibitemOpen
  \bibfield  {author} {\bibinfo {author} {\bibfnamefont {J.-F.}\ \bibnamefont
  {Li}}, \bibinfo {author} {\bibfnamefont {F.-Y.}\ \bibnamefont {Hou}},
  \bibinfo {author} {\bibfnamefont {Z.-F.}\ \bibnamefont {Cui}}, \bibinfo
  {author} {\bibfnamefont {H.-T.}\ \bibnamefont {Feng}}, \bibinfo {author}
  {\bibfnamefont {Y.}~\bibnamefont {Jiang}}, \ and\ \bibinfo {author}
  {\bibfnamefont {H.-S.}\ \bibnamefont {Zong}},\ }\href {\doibase
  10.1103/PhysRevD.90.073013} {\bibfield  {journal} {\bibinfo  {journal} {Phys.
  Rev. D}\ }\textbf {\bibinfo {volume} {90}},\ \bibinfo {pages} {073013}
  (\bibinfo {year} {2014})}\BibitemShut {NoStop}%
\bibitem [{\citenamefont {Klevansky}(1992)}]{RevModPhys.64.649}%
  \BibitemOpen
  \bibfield  {author} {\bibinfo {author} {\bibfnamefont {S.~P.}\ \bibnamefont
  {Klevansky}},\ }\href {\doibase 10.1103/RevModPhys.64.649} {\bibfield
  {journal} {\bibinfo  {journal} {Rev. Mod. Phys.}\ }\textbf {\bibinfo {volume}
  {64}},\ \bibinfo {pages} {649} (\bibinfo {year} {1992})}\BibitemShut
  {NoStop}%
\bibitem [{\citenamefont {Buballa}(2005)}]{Buballa2005205}%
  \BibitemOpen
  \bibfield  {author} {\bibinfo {author} {\bibfnamefont {M.}~\bibnamefont
  {Buballa}},\ }\href {\doibase
  http://dx.doi.org/10.1016/j.physrep.2004.11.004} {\bibfield  {journal}
  {\bibinfo  {journal} {Phys. Rep.}\ }\textbf {\bibinfo {volume} {407}},\
  \bibinfo {pages} {205 } (\bibinfo {year} {2005})}\BibitemShut {NoStop}%
\bibitem [{\citenamefont {Cui}\ \emph {et~al.}(2013)\citenamefont {Cui},
  \citenamefont {Shi}, \citenamefont {Xia}, \citenamefont {Jiang},\ and\
  \citenamefont {Zong}}]{Cui2013}%
  \BibitemOpen
  \bibfield  {author} {\bibinfo {author} {\bibfnamefont {Z.-F.}\ \bibnamefont
  {Cui}}, \bibinfo {author} {\bibfnamefont {C.}~\bibnamefont {Shi}}, \bibinfo
  {author} {\bibfnamefont {Y.-H.}\ \bibnamefont {Xia}}, \bibinfo {author}
  {\bibfnamefont {Y.}~\bibnamefont {Jiang}}, \ and\ \bibinfo {author}
  {\bibfnamefont {H.-S.}\ \bibnamefont {Zong}},\ }\href {\doibase
  10.1140/epjc/s10052-013-2612-6} {\bibfield  {journal} {\bibinfo  {journal}
  {Eur. Phys. J. C}\ }\textbf {\bibinfo {volume} {73}},\ \bibinfo {pages}
  {2612} (\bibinfo {year} {2013})}\BibitemShut {NoStop}%
\bibitem [{\citenamefont {Kohyama}\ \emph {et~al.}(2015)\citenamefont
  {Kohyama}, \citenamefont {Kimura},\ and\ \citenamefont
  {Inagaki}}]{KOHYAMA2015682}%
  \BibitemOpen
  \bibfield  {author} {\bibinfo {author} {\bibfnamefont {H.}~\bibnamefont
  {Kohyama}}, \bibinfo {author} {\bibfnamefont {D.}~\bibnamefont {Kimura}}, \
  and\ \bibinfo {author} {\bibfnamefont {T.}~\bibnamefont {Inagaki}},\ }\href
  {\doibase https://doi.org/10.1016/j.nuclphysb.2015.05.015} {\bibfield
  {journal} {\bibinfo  {journal} {Nucl. Phys. B}\ }\textbf {\bibinfo {volume}
  {896}},\ \bibinfo {pages} {682 } (\bibinfo {year} {2015})}\BibitemShut
  {NoStop}%
\bibitem [{\citenamefont {Fan}\ \emph {et~al.}(2017)\citenamefont {Fan},
  \citenamefont {Luo},\ and\ \citenamefont
  {Zong}}]{doi:10.1142/S0217751X17500610}%
  \BibitemOpen
  \bibfield  {author} {\bibinfo {author} {\bibfnamefont {W.}~\bibnamefont
  {Fan}}, \bibinfo {author} {\bibfnamefont {X.}~\bibnamefont {Luo}}, \ and\
  \bibinfo {author} {\bibfnamefont {H.-S.}\ \bibnamefont {Zong}},\ }\href
  {\doibase 10.1142/S0217751X17500610} {\bibfield  {journal} {\bibinfo
  {journal} {Int. J. Mod. Phys. A}\ }\textbf {\bibinfo {volume} {32}},\
  \bibinfo {pages} {1750061} (\bibinfo {year} {2017})}\BibitemShut {NoStop}%
\bibitem [{\citenamefont {Li}\ \emph {et~al.}(2019)\citenamefont {Li},
  \citenamefont {Yin},\ and\ \citenamefont {Zong}}]{PhysRevD.99.076006}%
  \BibitemOpen
  \bibfield  {author} {\bibinfo {author} {\bibfnamefont {C.-M.}\ \bibnamefont
  {Li}}, \bibinfo {author} {\bibfnamefont {P.-L.}\ \bibnamefont {Yin}}, \ and\
  \bibinfo {author} {\bibfnamefont {H.-S.}\ \bibnamefont {Zong}},\ }\href
  {\doibase 10.1103/PhysRevD.99.076006} {\bibfield  {journal} {\bibinfo
  {journal} {Phys. Rev. D}\ }\textbf {\bibinfo {volume} {99}},\ \bibinfo
  {pages} {076006} (\bibinfo {year} {2019})}\BibitemShut {NoStop}%
\bibitem [{\citenamefont {Chen}\ \emph {et~al.}(2011)\citenamefont {Chen},
  \citenamefont {Baldo}, \citenamefont {Burgio},\ and\ \citenamefont
  {Schulze}}]{PhysRevD.84.105023}%
  \BibitemOpen
  \bibfield  {author} {\bibinfo {author} {\bibfnamefont {H.}~\bibnamefont
  {Chen}}, \bibinfo {author} {\bibfnamefont {M.}~\bibnamefont {Baldo}},
  \bibinfo {author} {\bibfnamefont {G.~F.}\ \bibnamefont {Burgio}}, \ and\
  \bibinfo {author} {\bibfnamefont {H.-J.}\ \bibnamefont {Schulze}},\ }\href
  {\doibase 10.1103/PhysRevD.84.105023} {\bibfield  {journal} {\bibinfo
  {journal} {Phys. Rev. D}\ }\textbf {\bibinfo {volume} {84}},\ \bibinfo
  {pages} {105023} (\bibinfo {year} {2011})}\BibitemShut {NoStop}%
\bibitem [{\citenamefont {Chen}\ \emph {et~al.}(2015)\citenamefont {Chen},
  \citenamefont {Wei}, \citenamefont {Baldo}, \citenamefont {Burgio},\ and\
  \citenamefont {Schulze}}]{PhysRevD.91.105002}%
  \BibitemOpen
  \bibfield  {author} {\bibinfo {author} {\bibfnamefont {H.}~\bibnamefont
  {Chen}}, \bibinfo {author} {\bibfnamefont {J.-B.}\ \bibnamefont {Wei}},
  \bibinfo {author} {\bibfnamefont {M.}~\bibnamefont {Baldo}}, \bibinfo
  {author} {\bibfnamefont {G.~F.}\ \bibnamefont {Burgio}}, \ and\ \bibinfo
  {author} {\bibfnamefont {H.-J.}\ \bibnamefont {Schulze}},\ }\href {\doibase
  10.1103/PhysRevD.91.105002} {\bibfield  {journal} {\bibinfo  {journal} {Phys.
  Rev. D}\ }\textbf {\bibinfo {volume} {91}},\ \bibinfo {pages} {105002}
  (\bibinfo {year} {2015})}\BibitemShut {NoStop}%
\bibitem [{\citenamefont {Zhao}\ \emph {et~al.}(2015)\citenamefont {Zhao},
  \citenamefont {Xu}, \citenamefont {Yan}, \citenamefont {Luo}, \citenamefont
  {Liu},\ and\ \citenamefont {Zong}}]{PhysRevD.92.054012}%
  \BibitemOpen
  \bibfield  {author} {\bibinfo {author} {\bibfnamefont {T.}~\bibnamefont
  {Zhao}}, \bibinfo {author} {\bibfnamefont {S.-S.}\ \bibnamefont {Xu}},
  \bibinfo {author} {\bibfnamefont {Y.}~\bibnamefont {Yan}}, \bibinfo {author}
  {\bibfnamefont {X.-L.}\ \bibnamefont {Luo}}, \bibinfo {author} {\bibfnamefont
  {X.-J.}\ \bibnamefont {Liu}}, \ and\ \bibinfo {author} {\bibfnamefont
  {H.-S.}\ \bibnamefont {Zong}},\ }\href {\doibase 10.1103/PhysRevD.92.054012}
  {\bibfield  {journal} {\bibinfo  {journal} {Phys. Rev. D}\ }\textbf {\bibinfo
  {volume} {92}},\ \bibinfo {pages} {054012} (\bibinfo {year}
  {2015})}\BibitemShut {NoStop}%
\bibitem [{\citenamefont {Zhao}\ \emph {et~al.}(2017)\citenamefont {Zhao},
  \citenamefont {Li}, \citenamefont {Zhao}, \citenamefont {Yan}, \citenamefont
  {Luo},\ and\ \citenamefont {Zong}}]{doi:10.1142/S0217732317500511}%
  \BibitemOpen
  \bibfield  {author} {\bibinfo {author} {\bibfnamefont {T.}~\bibnamefont
  {Zhao}}, \bibinfo {author} {\bibfnamefont {C.-M.}\ \bibnamefont {Li}},
  \bibinfo {author} {\bibfnamefont {Y.-P.}\ \bibnamefont {Zhao}}, \bibinfo
  {author} {\bibfnamefont {Y.}~\bibnamefont {Yan}}, \bibinfo {author}
  {\bibfnamefont {X.-L.}\ \bibnamefont {Luo}}, \ and\ \bibinfo {author}
  {\bibfnamefont {H.-S.}\ \bibnamefont {Zong}},\ }\href {\doibase
  10.1142/S0217732317500511} {\bibfield  {journal} {\bibinfo  {journal} {Mod.
  Phys. Lett. A}\ }\textbf {\bibinfo {volume} {32}},\ \bibinfo {pages}
  {1750051} (\bibinfo {year} {2017})}\BibitemShut {NoStop}%
\bibitem [{\citenamefont {Wei}\ \emph {et~al.}(2017)\citenamefont {Wei},
  \citenamefont {Chen}, \citenamefont {Burgio},\ and\ \citenamefont
  {Schulze}}]{PhysRevD.96.043008}%
  \BibitemOpen
  \bibfield  {author} {\bibinfo {author} {\bibfnamefont {J.-B.}\ \bibnamefont
  {Wei}}, \bibinfo {author} {\bibfnamefont {H.}~\bibnamefont {Chen}}, \bibinfo
  {author} {\bibfnamefont {G.~F.}\ \bibnamefont {Burgio}}, \ and\ \bibinfo
  {author} {\bibfnamefont {H.-J.}\ \bibnamefont {Schulze}},\ }\href {\doibase
  10.1103/PhysRevD.96.043008} {\bibfield  {journal} {\bibinfo  {journal} {Phys.
  Rev. D}\ }\textbf {\bibinfo {volume} {96}},\ \bibinfo {pages} {043008}
  (\bibinfo {year} {2017})}\BibitemShut {NoStop}%
\bibitem [{\citenamefont {Bai}\ \emph {et~al.}(2018)\citenamefont {Bai},
  \citenamefont {Chen},\ and\ \citenamefont {Liu}}]{PhysRevD.97.023018}%
  \BibitemOpen
  \bibfield  {author} {\bibinfo {author} {\bibfnamefont {Z.}~\bibnamefont
  {Bai}}, \bibinfo {author} {\bibfnamefont {H.}~\bibnamefont {Chen}}, \ and\
  \bibinfo {author} {\bibfnamefont {Y.-x.}\ \bibnamefont {Liu}},\ }\href
  {\doibase 10.1103/PhysRevD.97.023018} {\bibfield  {journal} {\bibinfo
  {journal} {Phys. Rev. D}\ }\textbf {\bibinfo {volume} {97}},\ \bibinfo
  {pages} {023018} (\bibinfo {year} {2018})}\BibitemShut {NoStop}%
\bibitem [{\citenamefont {Blaschke}\ \emph {et~al.}(2010)\citenamefont
  {Blaschke}, \citenamefont {Berdermann},\ and\ \citenamefont
  {Lastowiecki}}]{10.1143/PTPS.186.81}%
  \BibitemOpen
  \bibfield  {author} {\bibinfo {author} {\bibfnamefont {D.}~\bibnamefont
  {Blaschke}}, \bibinfo {author} {\bibfnamefont {J.}~\bibnamefont
  {Berdermann}}, \ and\ \bibinfo {author} {\bibfnamefont {R.}~\bibnamefont
  {Lastowiecki}},\ }\href {\doibase 10.1143/PTPS.186.81} {\bibfield  {journal}
  {\bibinfo  {journal} {Prog. Theor. Phys. Suppl.}\ }\textbf {\bibinfo {volume}
  {186}},\ \bibinfo {pages} {81} (\bibinfo {year} {2010})}\BibitemShut
  {NoStop}%
\bibitem [{\citenamefont {Masuda}\ \emph {et~al.}(2013)\citenamefont {Masuda},
  \citenamefont {Hatsuda},\ and\ \citenamefont {Takatsuka}}]{Masuda01072013}%
  \BibitemOpen
  \bibfield  {author} {\bibinfo {author} {\bibfnamefont {K.}~\bibnamefont
  {Masuda}}, \bibinfo {author} {\bibfnamefont {T.}~\bibnamefont {Hatsuda}}, \
  and\ \bibinfo {author} {\bibfnamefont {T.}~\bibnamefont {Takatsuka}},\ }\href
  {\doibase 10.1093/ptep/ptt045} {\bibfield  {journal} {\bibinfo  {journal}
  {Prog. Theor. Exp. Phys.}\ }\textbf {\bibinfo {volume} {2013}} (\bibinfo
  {year} {2013}),\ 10.1093/ptep/ptt045}\BibitemShut {NoStop}%
\bibitem [{\citenamefont {Whittenbury}\ \emph {et~al.}(2016)\citenamefont
  {Whittenbury}, \citenamefont {Matevosyan},\ and\ \citenamefont
  {Thomas}}]{PhysRevC.93.035807}%
  \BibitemOpen
  \bibfield  {author} {\bibinfo {author} {\bibfnamefont {D.~L.}\ \bibnamefont
  {Whittenbury}}, \bibinfo {author} {\bibfnamefont {H.~H.}\ \bibnamefont
  {Matevosyan}}, \ and\ \bibinfo {author} {\bibfnamefont {A.~W.}\ \bibnamefont
  {Thomas}},\ }\href {\doibase 10.1103/PhysRevC.93.035807} {\bibfield
  {journal} {\bibinfo  {journal} {Phys. Rev. C}\ }\textbf {\bibinfo {volume}
  {93}},\ \bibinfo {pages} {035807} (\bibinfo {year} {2016})}\BibitemShut
  {NoStop}%
\bibitem [{\citenamefont {Pereira}\ \emph {et~al.}(2016)\citenamefont
  {Pereira}, \citenamefont {Costa},\ and\ \citenamefont
  {Providencia}}]{PhysRevD.94.094001}%
  \BibitemOpen
  \bibfield  {author} {\bibinfo {author} {\bibfnamefont {R.~C.}\ \bibnamefont
  {Pereira}}, \bibinfo {author} {\bibfnamefont {P.}~\bibnamefont {Costa}}, \
  and\ \bibinfo {author} {\bibfnamefont {C.}~\bibnamefont {Providencia}},\
  }\href {\doibase 10.1103/PhysRevD.94.094001} {\bibfield  {journal} {\bibinfo
  {journal} {Phys. Rev. D}\ }\textbf {\bibinfo {volume} {94}},\ \bibinfo
  {pages} {094001} (\bibinfo {year} {2016})}\BibitemShut {NoStop}%
\bibitem [{\citenamefont {Li}\ \emph {et~al.}(2017)\citenamefont {Li},
  \citenamefont {Zhang}, \citenamefont {Zhao}, \citenamefont {Zhao},\ and\
  \citenamefont {Zong}}]{PhysRevD.95.056018}%
  \BibitemOpen
  \bibfield  {author} {\bibinfo {author} {\bibfnamefont {C.-M.}\ \bibnamefont
  {Li}}, \bibinfo {author} {\bibfnamefont {J.-L.}\ \bibnamefont {Zhang}},
  \bibinfo {author} {\bibfnamefont {T.}~\bibnamefont {Zhao}}, \bibinfo {author}
  {\bibfnamefont {Y.-P.}\ \bibnamefont {Zhao}}, \ and\ \bibinfo {author}
  {\bibfnamefont {H.-S.}\ \bibnamefont {Zong}},\ }\href {\doibase
  10.1103/PhysRevD.95.056018} {\bibfield  {journal} {\bibinfo  {journal} {Phys.
  Rev. D}\ }\textbf {\bibinfo {volume} {95}},\ \bibinfo {pages} {056018}
  (\bibinfo {year} {2017})}\BibitemShut {NoStop}%
\bibitem [{\citenamefont {Li}\ \emph {et~al.}(2018{\natexlab{a}})\citenamefont
  {Li}, \citenamefont {Zhang}, \citenamefont {Yan}, \citenamefont {Huang},\
  and\ \citenamefont {Zong}}]{PhysRevD.97.103013}%
  \BibitemOpen
  \bibfield  {author} {\bibinfo {author} {\bibfnamefont {C.-M.}\ \bibnamefont
  {Li}}, \bibinfo {author} {\bibfnamefont {J.-L.}\ \bibnamefont {Zhang}},
  \bibinfo {author} {\bibfnamefont {Y.}~\bibnamefont {Yan}}, \bibinfo {author}
  {\bibfnamefont {Y.-F.}\ \bibnamefont {Huang}}, \ and\ \bibinfo {author}
  {\bibfnamefont {H.-S.}\ \bibnamefont {Zong}},\ }\href {\doibase
  10.1103/PhysRevD.97.103013} {\bibfield  {journal} {\bibinfo  {journal} {Phys.
  Rev. D}\ }\textbf {\bibinfo {volume} {97}},\ \bibinfo {pages} {103013}
  (\bibinfo {year} {2018}{\natexlab{a}})}\BibitemShut {NoStop}%
\bibitem [{\citenamefont {Li}\ \emph {et~al.}(2018{\natexlab{b}})\citenamefont
  {Li}, \citenamefont {Yan}, \citenamefont {Geng}, \citenamefont {Huang},\ and\
  \citenamefont {Zong}}]{PhysRevD.98.083013}%
  \BibitemOpen
  \bibfield  {author} {\bibinfo {author} {\bibfnamefont {C.-M.}\ \bibnamefont
  {Li}}, \bibinfo {author} {\bibfnamefont {Y.}~\bibnamefont {Yan}}, \bibinfo
  {author} {\bibfnamefont {J.-J.}\ \bibnamefont {Geng}}, \bibinfo {author}
  {\bibfnamefont {Y.-F.}\ \bibnamefont {Huang}}, \ and\ \bibinfo {author}
  {\bibfnamefont {H.-S.}\ \bibnamefont {Zong}},\ }\href {\doibase
  10.1103/PhysRevD.98.083013} {\bibfield  {journal} {\bibinfo  {journal} {Phys.
  Rev. D}\ }\textbf {\bibinfo {volume} {98}},\ \bibinfo {pages} {083013}
  (\bibinfo {year} {2018}{\natexlab{b}})}\BibitemShut {NoStop}%
\bibitem [{\citenamefont {Wang}\ \emph
  {et~al.}(2019{\natexlab{b}})\citenamefont {Wang}, \citenamefont {Cao},\ and\
  \citenamefont {Zong}}]{Wang2019}%
  \BibitemOpen
  \bibfield  {author} {\bibinfo {author} {\bibfnamefont {F.}~\bibnamefont
  {Wang}}, \bibinfo {author} {\bibfnamefont {Y.}~\bibnamefont {Cao}}, \ and\
  \bibinfo {author} {\bibfnamefont {H.}~\bibnamefont {Zong}},\ }\href {\doibase
  10.1088/1674-1137/43/8/084102} {\bibfield  {journal} {\bibinfo  {journal}
  {Chin. Phys. C}\ }\textbf {\bibinfo {volume} {43}},\ \bibinfo {pages}
  {084102} (\bibinfo {year} {2019}{\natexlab{b}})}\BibitemShut {NoStop}%
\bibitem [{\citenamefont {Klimt}\ \emph
  {et~al.}(1990{\natexlab{a}})\citenamefont {Klimt}, \citenamefont {Lutz},\
  and\ \citenamefont {Weise}}]{KLIMT1990386}%
  \BibitemOpen
  \bibfield  {author} {\bibinfo {author} {\bibfnamefont {S.}~\bibnamefont
  {Klimt}}, \bibinfo {author} {\bibfnamefont {M.}~\bibnamefont {Lutz}}, \ and\
  \bibinfo {author} {\bibfnamefont {W.}~\bibnamefont {Weise}},\ }\href
  {\doibase https://doi.org/10.1016/0370-2693(90)91003-T} {\bibfield  {journal}
  {\bibinfo  {journal} {Phys. Lett. B}\ }\textbf {\bibinfo {volume} {249}},\
  \bibinfo {pages} {386 } (\bibinfo {year} {1990}{\natexlab{a}})}\BibitemShut
  {NoStop}%
\bibitem [{\citenamefont {Klimt}\ \emph
  {et~al.}(1990{\natexlab{b}})\citenamefont {Klimt}, \citenamefont {Lutz},
  \citenamefont {Vogl},\ and\ \citenamefont {Weise}}]{KLIMT1990429}%
  \BibitemOpen
  \bibfield  {author} {\bibinfo {author} {\bibfnamefont {S.}~\bibnamefont
  {Klimt}}, \bibinfo {author} {\bibfnamefont {M.}~\bibnamefont {Lutz}},
  \bibinfo {author} {\bibfnamefont {U.}~\bibnamefont {Vogl}}, \ and\ \bibinfo
  {author} {\bibfnamefont {W.}~\bibnamefont {Weise}},\ }\href {\doibase
  https://doi.org/10.1016/0375-9474(90)90123-4} {\bibfield  {journal} {\bibinfo
   {journal} {Nucl. Phys. A}\ }\textbf {\bibinfo {volume} {516}},\ \bibinfo
  {pages} {429 } (\bibinfo {year} {1990}{\natexlab{b}})}\BibitemShut {NoStop}%
\bibitem [{\citenamefont {Bernard}\ \emph {et~al.}(1988)\citenamefont
  {Bernard}, \citenamefont {Jaffe},\ and\ \citenamefont
  {Meissner}}]{BERNARD1988753}%
  \BibitemOpen
  \bibfield  {author} {\bibinfo {author} {\bibfnamefont {V.}~\bibnamefont
  {Bernard}}, \bibinfo {author} {\bibfnamefont {R.}~\bibnamefont {Jaffe}}, \
  and\ \bibinfo {author} {\bibfnamefont {U.-G.}\ \bibnamefont {Meissner}},\
  }\href {\doibase https://doi.org/10.1016/0550-3213(88)90127-7} {\bibfield
  {journal} {\bibinfo  {journal} {Nucl. Phys. B}\ }\textbf {\bibinfo {volume}
  {308}},\ \bibinfo {pages} {753 } (\bibinfo {year} {1988})}\BibitemShut
  {NoStop}%
\bibitem [{\citenamefont {Hatsuda}\ and\ \citenamefont
  {Kunihiro}(1985)}]{10.1143/PTP.74.765}%
  \BibitemOpen
  \bibfield  {author} {\bibinfo {author} {\bibfnamefont {T.}~\bibnamefont
  {Hatsuda}}\ and\ \bibinfo {author} {\bibfnamefont {T.}~\bibnamefont
  {Kunihiro}},\ }\href {\doibase 10.1143/PTP.74.765} {\bibfield  {journal}
  {\bibinfo  {journal} {Prog. Theor. Phys.}\ }\textbf {\bibinfo {volume}
  {74}},\ \bibinfo {pages} {765} (\bibinfo {year} {1985})}\BibitemShut
  {NoStop}%
\bibitem [{\citenamefont {Zong}\ \emph {et~al.}(2005)\citenamefont {Zong},
  \citenamefont {Chang}, \citenamefont {Hou}, \citenamefont {Sun},\ and\
  \citenamefont {Liu}}]{PhysRevC.71.015205}%
  \BibitemOpen
  \bibfield  {author} {\bibinfo {author} {\bibfnamefont {H.-S.}\ \bibnamefont
  {Zong}}, \bibinfo {author} {\bibfnamefont {L.}~\bibnamefont {Chang}},
  \bibinfo {author} {\bibfnamefont {F.-Y.}\ \bibnamefont {Hou}}, \bibinfo
  {author} {\bibfnamefont {W.-M.}\ \bibnamefont {Sun}}, \ and\ \bibinfo
  {author} {\bibfnamefont {Y.-X.}\ \bibnamefont {Liu}},\ }\href {\doibase
  10.1103/PhysRevC.71.015205} {\bibfield  {journal} {\bibinfo  {journal} {Phys.
  Rev. C}\ }\textbf {\bibinfo {volume} {71}},\ \bibinfo {pages} {015205}
  (\bibinfo {year} {2005})}\BibitemShut {NoStop}%
\bibitem [{\citenamefont {Tanabashi}\ and\ \citenamefont
  {et~al.}(2018)}]{PhysRevD.98.030001}%
  \BibitemOpen
  \bibfield  {author} {\bibinfo {author} {\bibfnamefont {M.}~\bibnamefont
  {Tanabashi}}\ and\ \bibinfo {author} {\bibnamefont {et~al.}} (\bibinfo
  {collaboration} {Particle Data Group}),\ }\href {\doibase
  10.1103/PhysRevD.98.030001} {\bibfield  {journal} {\bibinfo  {journal} {Phys.
  Rev. D}\ }\textbf {\bibinfo {volume} {98}},\ \bibinfo {pages} {030001}
  (\bibinfo {year} {2018})}\BibitemShut {NoStop}%
\bibitem [{\citenamefont {Zong}\ and\ \citenamefont
  {Sun}(2008{\natexlab{a}})}]{doi:10.1142/S0217751X08040457}%
  \BibitemOpen
  \bibfield  {author} {\bibinfo {author} {\bibfnamefont {H.-S.}\ \bibnamefont
  {Zong}}\ and\ \bibinfo {author} {\bibfnamefont {W.-M.}\ \bibnamefont {Sun}},\
  }\href {\doibase 10.1142/S0217751X08040457} {\bibfield  {journal} {\bibinfo
  {journal} {Int. J. Mod. Phys. A}\ }\textbf {\bibinfo {volume} {23}},\
  \bibinfo {pages} {3591} (\bibinfo {year} {2008}{\natexlab{a}})}\BibitemShut
  {NoStop}%
\bibitem [{\citenamefont {Zong}\ and\ \citenamefont
  {Sun}(2008{\natexlab{b}})}]{PhysRevD.78.054001}%
  \BibitemOpen
  \bibfield  {author} {\bibinfo {author} {\bibfnamefont {H.-S.}\ \bibnamefont
  {Zong}}\ and\ \bibinfo {author} {\bibfnamefont {W.-M.}\ \bibnamefont {Sun}},\
  }\href {\doibase 10.1103/PhysRevD.78.054001} {\bibfield  {journal} {\bibinfo
  {journal} {Phys. Rev. D}\ }\textbf {\bibinfo {volume} {78}},\ \bibinfo
  {pages} {054001} (\bibinfo {year} {2008}{\natexlab{b}})}\BibitemShut
  {NoStop}%
\bibitem [{\citenamefont {Chernodub}\ \emph {et~al.}(2018)\citenamefont
  {Chernodub}, \citenamefont {Goy}, \citenamefont {Molochkov},\ and\
  \citenamefont {Nguyen}}]{PhysRevLett.121.191601}%
  \BibitemOpen
  \bibfield  {author} {\bibinfo {author} {\bibfnamefont {M.~N.}\ \bibnamefont
  {Chernodub}}, \bibinfo {author} {\bibfnamefont {V.~A.}\ \bibnamefont {Goy}},
  \bibinfo {author} {\bibfnamefont {A.~V.}\ \bibnamefont {Molochkov}}, \ and\
  \bibinfo {author} {\bibfnamefont {H.~H.}\ \bibnamefont {Nguyen}},\ }\href
  {\doibase 10.1103/PhysRevLett.121.191601} {\bibfield  {journal} {\bibinfo
  {journal} {Phys. Rev. Lett.}\ }\textbf {\bibinfo {volume} {121}},\ \bibinfo
  {pages} {191601} (\bibinfo {year} {2018})}\BibitemShut {NoStop}%
\bibitem [{\citenamefont {Chernodub}\ \emph {et~al.}(2019)\citenamefont
  {Chernodub}, \citenamefont {Goy},\ and\ \citenamefont
  {Molochkov}}]{Chernodub:2019nct}%
  \BibitemOpen
  \bibfield  {author} {\bibinfo {author} {\bibfnamefont {M.~N.}\ \bibnamefont
  {Chernodub}}, \bibinfo {author} {\bibfnamefont {V.~A.}\ \bibnamefont {Goy}},
  \ and\ \bibinfo {author} {\bibfnamefont {A.~V.}\ \bibnamefont {Molochkov}},\
  }\bibfield  {booktitle} {\emph {\bibinfo {booktitle} {{Proceedings, 13th
  Conference on Quark Confinement and the Hadron Spectrum (Confinement XIII):
  Maynooth, Ireland}}},\ }\href {\doibase 10.22323/1.336.0006} {\bibfield
  {journal} {\bibinfo  {journal} {PoS}\ }\textbf {\bibinfo {volume}
  {Confinement2018}},\ \bibinfo {pages} {006} (\bibinfo {year} {2019})},\
  \Eprint {http://arxiv.org/abs/1901.04754} {arXiv:1901.04754 [hep-th]}
  \BibitemShut {NoStop}%
\bibitem [{\citenamefont {Xu}\ \emph {et~al.}(2018)\citenamefont {Xu},
  \citenamefont {Cui}, \citenamefont {Sun},\ and\ \citenamefont
  {Zong}}]{0954-3899-45-10-105001}%
  \BibitemOpen
  \bibfield  {author} {\bibinfo {author} {\bibfnamefont {S.-S.}\ \bibnamefont
  {Xu}}, \bibinfo {author} {\bibfnamefont {Z.-F.}\ \bibnamefont {Cui}},
  \bibinfo {author} {\bibfnamefont {A.}~\bibnamefont {Sun}}, \ and\ \bibinfo
  {author} {\bibfnamefont {H.-S.}\ \bibnamefont {Zong}},\ }\href
  {http://stacks.iop.org/0954-3899/45/i=10/a=105001} {\bibfield  {journal}
  {\bibinfo  {journal} {J. Phys. G: Nucl. Part. Phys.}\ }\textbf {\bibinfo
  {volume} {45}},\ \bibinfo {pages} {105001} (\bibinfo {year}
  {2018})}\BibitemShut {NoStop}%
\bibitem [{\citenamefont {Song}\ \emph {et~al.}(1992)\citenamefont {Song},
  \citenamefont {Enke},\ and\ \citenamefont {Jiarong}}]{PhysRevD.46.3211}%
  \BibitemOpen
  \bibfield  {author} {\bibinfo {author} {\bibfnamefont {G.}~\bibnamefont
  {Song}}, \bibinfo {author} {\bibfnamefont {W.}~\bibnamefont {Enke}}, \ and\
  \bibinfo {author} {\bibfnamefont {L.}~\bibnamefont {Jiarong}},\ }\href
  {\doibase 10.1103/PhysRevD.46.3211} {\bibfield  {journal} {\bibinfo
  {journal} {Phys. Rev. D}\ }\textbf {\bibinfo {volume} {46}},\ \bibinfo
  {pages} {3211} (\bibinfo {year} {1992})}\BibitemShut {NoStop}%
\bibitem [{\citenamefont {Lu}\ \emph {et~al.}(1998)\citenamefont {Lu},
  \citenamefont {Tsushima}, \citenamefont {Thomas}, \citenamefont {Williams},\
  and\ \citenamefont {Saito}}]{LU1998443}%
  \BibitemOpen
  \bibfield  {author} {\bibinfo {author} {\bibfnamefont {D.}~\bibnamefont
  {Lu}}, \bibinfo {author} {\bibfnamefont {K.}~\bibnamefont {Tsushima}},
  \bibinfo {author} {\bibfnamefont {A.}~\bibnamefont {Thomas}}, \bibinfo
  {author} {\bibfnamefont {A.}~\bibnamefont {Williams}}, \ and\ \bibinfo
  {author} {\bibfnamefont {K.}~\bibnamefont {Saito}},\ }\href {\doibase
  https://doi.org/10.1016/S0375-9474(98)00181-X} {\bibfield  {journal}
  {\bibinfo  {journal} {Nucl. Phys. A}\ }\textbf {\bibinfo {volume} {634}},\
  \bibinfo {pages} {443 } (\bibinfo {year} {1998})}\BibitemShut {NoStop}%
\bibitem [{\citenamefont {Zhou}\ \emph {et~al.}(2018)\citenamefont {Zhou},
  \citenamefont {Zhou},\ and\ \citenamefont {Li}}]{PhysRevD.97.083015}%
  \BibitemOpen
  \bibfield  {author} {\bibinfo {author} {\bibfnamefont {E.-P.}\ \bibnamefont
  {Zhou}}, \bibinfo {author} {\bibfnamefont {X.}~\bibnamefont {Zhou}}, \ and\
  \bibinfo {author} {\bibfnamefont {A.}~\bibnamefont {Li}},\ }\href {\doibase
  10.1103/PhysRevD.97.083015} {\bibfield  {journal} {\bibinfo  {journal} {Phys.
  Rev. D}\ }\textbf {\bibinfo {volume} {97}},\ \bibinfo {pages} {083015}
  (\bibinfo {year} {2018})}\BibitemShut {NoStop}%
\bibitem [{\citenamefont {Yan}\ \emph {et~al.}(2012)\citenamefont {Yan},
  \citenamefont {Cao}, \citenamefont {Luo}, \citenamefont {Sun},\ and\
  \citenamefont {Zong}}]{PhysRevD.86.114028}%
  \BibitemOpen
  \bibfield  {author} {\bibinfo {author} {\bibfnamefont {Y.}~\bibnamefont
  {Yan}}, \bibinfo {author} {\bibfnamefont {J.}~\bibnamefont {Cao}}, \bibinfo
  {author} {\bibfnamefont {X.-L.}\ \bibnamefont {Luo}}, \bibinfo {author}
  {\bibfnamefont {W.-M.}\ \bibnamefont {Sun}}, \ and\ \bibinfo {author}
  {\bibfnamefont {H.-S.}\ \bibnamefont {Zong}},\ }\href {\doibase
  10.1103/PhysRevD.86.114028} {\bibfield  {journal} {\bibinfo  {journal} {Phys.
  Rev. D}\ }\textbf {\bibinfo {volume} {86}},\ \bibinfo {pages} {114028}
  (\bibinfo {year} {2012})}\BibitemShut {NoStop}%
\bibitem [{\citenamefont {Benvenuto}\ and\ \citenamefont
  {Lugones}(1995)}]{PhysRevD.51.1989}%
  \BibitemOpen
  \bibfield  {author} {\bibinfo {author} {\bibfnamefont {O.~G.}\ \bibnamefont
  {Benvenuto}}\ and\ \bibinfo {author} {\bibfnamefont {G.}~\bibnamefont
  {Lugones}},\ }\href {\doibase 10.1103/PhysRevD.51.1989} {\bibfield  {journal}
  {\bibinfo  {journal} {Phys. Rev. D}\ }\textbf {\bibinfo {volume} {51}},\
  \bibinfo {pages} {1989} (\bibinfo {year} {1995})}\BibitemShut {NoStop}%
\bibitem [{\citenamefont {Hinderer}\ \emph {et~al.}(2010)\citenamefont
  {Hinderer}, \citenamefont {Lackey}, \citenamefont {Lang},\ and\ \citenamefont
  {Read}}]{PhysRevD.81.123016}%
  \BibitemOpen
  \bibfield  {author} {\bibinfo {author} {\bibfnamefont {T.}~\bibnamefont
  {Hinderer}}, \bibinfo {author} {\bibfnamefont {B.~D.}\ \bibnamefont
  {Lackey}}, \bibinfo {author} {\bibfnamefont {R.~N.}\ \bibnamefont {Lang}}, \
  and\ \bibinfo {author} {\bibfnamefont {J.~S.}\ \bibnamefont {Read}},\ }\href
  {\doibase 10.1103/PhysRevD.81.123016} {\bibfield  {journal} {\bibinfo
  {journal} {Phys. Rev. D}\ }\textbf {\bibinfo {volume} {81}},\ \bibinfo
  {pages} {123016} (\bibinfo {year} {2010})}\BibitemShut {NoStop}%
\end{thebibliography}%
\end{document}